\documentclass[a4paper,11pt]{article}
\usepackage{jcappub}
\usepackage{amsmath}
\usepackage{graphicx}
\usepackage{latexsym}
\usepackage{xspace}
\usepackage{color}
\usepackage{hyperref} 
\usepackage{bm}
\usepackage{relsize}

\makeatletter
\gdef\@fpheader{}
\makeatother

\newcommand{\ie}{i.e.\xspace}
\newcommand{\eg}{e.g.\xspace}
\newcommand{\etc}{etc.}

\newcommand{\mean}[1]{\left\langle #1 \right\rangle}

\newcommand{\dd}{\mathrm{d}}
\newcommand{\ee}{e}

\newcommand{\sss}[1]{{\scriptscriptstyle{#1}}}

\newcommand{\uPl}{\mathrm{Pl}}
\newcommand{\uin}{\mathrm{in}}

\newcommand{\umax}{\mathrm{max}}
\newcommand{\uend}{\mathrm{end}}

\newcommand{\ucl}{\mathrm{cl}}

\newcommand{\uc}{\mathrm{c}}

\newcommand{\uS}{\mathrm{S}}

\newcommand{\usssS}{\sss{\uS}}

\newcommand{\usssPl}{\sss{\uPl}}

\newcommand{\nS}{n_\usssS}

\newcommand{\uNL}{\mathrm{NL}}

\newcommand{\calP}{\mathcal{P}}

\newcommand{\Mp}{M_\usssPl}

\newcommand{\fnl}{f_\uNL}

\newcommand{\efolds}{$e$-folds~}
\newcommand{\efold}{$e$-fold}

\newcommand{\beq}{\begin{equation}}
\newcommand{\eeq}{\end{equation}}
\newcommand{\bea}{\begin{eqnarray}}
\newcommand{\eea}{\end{eqnarray}}

\newlength{\wsingfig}
\newlength{\wdblefig}
\newlength{\wquadfig}
\newlength{\wtriplefig}
\newcommand{\Eq}[1]{Eq.~(\ref{#1})}
\newcommand{\Eqs}[1]{Eqs.~(\ref{#1})}
\newcommand{\Fig}[1]{Fig.~{\ref{#1}}}
\newcommand{\Figs}[1]{Figs.~{\ref{#1}}}
\newcommand{\Ref}[1]{Ref.~{\cite{#1}}}
\newcommand{\Refs}[1]{Refs.~{\cite{#1}}}

\subheader{}

\title{Correlation Functions\\ in Stochastic Inflation}

\author[a]{Vincent Vennin}
\author[b,c]{and Alexei A.\ Starobinsky}

\affiliation[a]{Institute of Cosmology \& Gravitation, University of Portsmouth, Dennis Sciama Building, Burnaby Road, Portsmouth, PO1 3FX, United Kingdom}
\affiliation[b]{L. D. Landau Institute for Theoretical Physics RAS,
Moscow 119334, Russian Federation}
\affiliation[c]{Department of Physics and Astronomy, Institute for Theoretical Physics, Utrecht University, 3508 TD Utrecht, The Netherlands}

\emailAdd{vincent.vennin@port.ac.uk}
\emailAdd{alstar@landau.ac.ru}

\date{today}

\begin{document}

\abstract{
Combining the stochastic and $\delta N$ formalisms, we derive non-perturbative analytical expressions for all correlation functions of scalar perturbations in single-field, slow-roll inflation. The standard, classical formulas are recovered as saddle-point limits of the full results. This yields a classicality criterion that shows that stochastic effects are small only if the potential is sub-Planckian \emph{and} not too flat. The saddle-point approximation also provides an expansion scheme for calculating stochastic corrections to observable quantities perturbatively in this regime. In the opposite regime, we show that a strong suppression in the power spectrum is generically obtained, and we comment on the physical implications of this effect. 
}


\arxivnumber{1506.04732}

\maketitle
\section{Introduction}
Inflation is one of the leading paradigms describing the physical conditions that prevailed in the very early Universe~\cite{Starobinsky:1980te, Sato:1980yn, Guth:1980zm, Linde:1981mu, Albrecht:1982wi, Linde:1983gd}. It is a phase of accelerated expansion that solves the puzzles of the standard hot Big Bang model, and it provides a causal mechanism for generating scalar~\cite{Mukhanov:1981xt, Hawking:1982cz,  Starobinsky:1982ee, Guth:1982ec, Bardeen:1983qw} and tensor~\cite{Starobinsky:1979ty} inhomogeneous perturbations on cosmological scales. These inhomogeneities result from the parametric amplification of the  vacuum quantum fluctuations of the gravitational and matter fields during the accelerated expansion. 

The transition from these quantum fluctuations to classical but stochastic density perturbations~\cite{Polarski:1995jg,Lesgourgues:1996jc,Kiefer:2008ku,Martin:2012pea} gives rise to the stochastic inflation formalism~\cite{Starobinsky:1986fx, Nambu:1987ef, Nambu:1988je, Kandrup:1988sc, Nakao:1988yi, Nambu:1989uf, Mollerach:1990zf, Linde:1993xx, Starobinsky:1994bd}.\footnote{This formalism was, in fact, first used in \Ref{Starobinsky:1982ee} at the level of the Langevin equation, from which results lying beyond the one-loop approximation for the inflaton field were obtained.} It consists of an effective theory for the long-wavelength parts of the quantum fields, which are ``coarse grained'' at a fixed physical scale (\ie non-expanding), somewhat larger than the Hubble radius during the whole inflationary period.\footnote{More precisely, the coarse grained part of the field consists of the modes $k$ for which $k\lesssim \sigma a H$. Here, $\sigma$ is a cutoff parameter satisfying~\cite{Starobinsky:1994bd} $e^{-1/(3\epsilon_1)}\ll \sigma\ll 1$, where $\epsilon_1$ is the first slow-roll parameter. Under this condition, the physical results are independent of $\sigma$.} The non-commutative parts of this coarse grained field $\varphi$ are small, and at this scale, short-wavelength quantum fluctuations have negligible non-commutative parts too. In this framework, they act as a classical noise on the dynamics of the super-Hubble scales, and $\varphi$ can thus be described by a stochastic classical theory, following the Langevin equation
\beq
\label{eq:Langevin}
\frac{\dd \varphi}{\dd N}=-\frac{V^\prime}{3H^2}+\frac{H}{2\pi}\xi\left(N\right)\, .
\eeq
This equation is valid at leading order in slow roll. Time is labeled by the number of \efolds $N\equiv\ln a$, where $a$ is the scale factor. The Hubble parameter $H\equiv \dd a/(a \dd t)$ is related to the potential $V$ via the slow-roll Friedmann equation $H^2\simeq V/(3\Mp^2)$, where $\Mp$ is the reduced Planck mass. The dynamics of $\varphi$ is then driven by two terms. The first one, proportional to $V^\prime$ (where a prime denotes a derivative with respect to the inflaton field), is the classical drift. The second one involves a white Gaussian noise, $\xi$, and renders the dynamics stochastic. It is such that $\left\langle\xi\left(N\right)\right\rangle=0$ and $\left\langle\xi\left(N\right)\xi\left(N^\prime\right)\right\rangle=\delta\left(N-N^\prime\right)$. 

The stochastic formalism thus accounts for the quantum modification of the super-Hubble scales dynamics. It allows us to calculate quantum corrections on background quantities beyond the one-loop approximation for the inflaton scalar field $\phi$ (in fact, beyond any finite number of inflaton loops)  and to calculate such quantities as \eg the probability distribution and any moments of the number of inflationary \efolds in a given point. In turn, cosmological perturbations are affected too, and a natural question to address within the stochastic framework is therefore how quantum effects modify inflationary observable predictions. This is the main motivation of the present work.

Stochastic inflation is a powerful tool for calculating correlation functions of quantum fields during inflation. In \Refs{Starobinsky:1994bd, Finelli:2008zg, Finelli:2010sh, Garbrecht:2013coa, Garbrecht:2014dca}, it is shown that standard results of quantum field theory (QFT) are recovered by the stochastic formalism for test scalar fields on fixed inflationary backgrounds, for any finite number of scalar loops and potentially beyond. This result is even extended to scalar electrodynamics during inflation in \Refs{Prokopec:2007ak, Prokopec:2008gw} and to derivative interactions and constrained fields in \Ref{Tsamis:2005hd}.  In \Ref{Finelli:2008zg}, fluctuations of a non-test inflaton field have been studied, too. In this last case, the calculation is performed at linear order in the noise, that is, assuming that the distribution of the coarse grained field remains peaked around its classical value $\phi_\uc$, where $\phi_\uc$ is the solution of \Eq{eq:Langevin} without the noise term. However, it may happen that the quantum kicks dominate over the classical drift and provide the main contribution to the inflationary dynamics in some flat parts of the potential. It is therefore legitimate to wonder what observable imprints could be left in such cases. In order to deal with observable quantities, the goal of this paper is therefore to calculate the correlation functions of inflationary perturbations in full generality, taking backreaction of created inflaton fluctuations on its background value into account, starting from \Eq{eq:Langevin} and without relying on a perturbative expansion in the noise.\footnote{In this connection, the approach of \Ref{Enqvist:2008kt} is close to ours. However, we use a different form of the Fokker-Planck equation, a different initial condition for the inflaton probability distribution, and a different form of the $\delta N$ formalism (which, in fact, may be called $N$ formalism) that does not use an expansion in $\delta N$ and in the metric perturbation $\zeta$ (in fact, these two quantities are not small in the so called regime of eternal inflation).}

This work is organized as follows. In section~\ref{sec:statDistrib}, we first discuss the issue of the time variable choice in the Langevin equation (we further elaborate on this aspect in appendix~\ref{sec:app:whyN}). This allows us to set a few notations, and to already argue why some of the effects later obtained (but not all) are Planck suppressed. In section~\ref{sec:PSinstocha}, we turn to the calculation of the correlation functions of primordial cosmological perturbations, without assuming them to be small. We first review different methods that have been used in the literature, and motivate our choice of combining the stochastic and $\delta N$ formalisms. We then settle our computational strategy and proceed with the calculation itself. Results are presented in section~\ref{sec:Results}; see in particular \Eqs{eq:PS:fullstocha} and~(\ref{eq:fnl:exact}). We show that the standard formulas are recovered in a ``classical'' limit that we carefully define, and discuss the regimes where they are not valid. Finally, in section~\ref{sec:Conclusion}, we summarize our main results and conclude.
\section{Time Variable Issue}
\label{sec:statDistrib}
Because of the Friedmann equation, the Hubble parameter appearing in \Eq{eq:Langevin} is sourced by the inflaton field itself, through the slow-roll function $H(\varphi)$. At leading order in the noise, one simply has $H(\varphi)=H(\phi_\uc)$, which is a classical quantity. Beyond the leading order, however, $H$ is dependent on the full coarse grained field and is therefore a stochastic quantity.\footnote{Hereafter, by ``stochastic quantities'', we simply refer to realization dependent quantities, as opposed to quantities that are fixed for all realizations.} This has two consequences. The first one is that starting from a classical time label, any other time variable defined through $a$ or $H$ is a stochastic quantity, and cannot be used to label the Langevin equation, otherwise one would describe a physically different process. The time label must therefore be carefully specified. The second one is that, since $H$ is related to the curvature of space-time, its stochasticity has to do with the one of space-time itself. We are thus a priori describing effective quantum gravitational effects, corresponding to the gravitational- and self-interactions of the inflaton field. The corresponding corrections should therefore remain small as long as the energy density of the inflaton field is small compared to the Planck scale. For this reason, it is convenient to define the dimensionless potential
\beq
\label{eq:v:def}
v=\frac{V}{24\pi^2\Mp^4}\approx \frac{H^2}{8\pi^2 \Mp^2}
\eeq
which we will make use of extensively in the following. Before turning to the calculation of the correlation functions, in this section, we show on a simple example why different time labels in the Langevin equation typically yield results that differ by $\propto v$ corrections. 

First, let us recast the stochastic process (\ref{eq:Langevin}) through a Fokker-Planck equation, which governs the time evolution of the probability density $P(\phi,N)$ that $\varphi=\phi$ at time $N$. In the It\^o interpretation\footnote{More generally, the last term in \Eq{eq:FokkerPlanck} can be written in the form
\beq
\frac{\partial}{\partial \phi}
\left\lbrace\frac{H^{2\alpha}}{8\pi^2}\frac{\partial}{\partial\phi}\left[H^{2(1-\alpha)}P\left(\phi,N\right)\right]\right\rbrace
\eeq
with $0\le \alpha\le 1$, where $\alpha=0$ corresponds to the It\^o interpretation and $\alpha=1/2$ to the Stratonovich one~\cite{Stratonovich:1966}. However, analysis shows that keeping terms explicitly depending on $\alpha$ exceeds the accuracy of the stochastic approach in its leading approximation~(\ref{eq:Langevin}). In particular, corrections to the noise term due to self-interactions of small-scale fluctuations (if they exist) are at least of the same order or even larger.}~\cite{Starobinsky:1986fx,Winitzki:1995pg, Vilenkin:1999kd}, it reads\footnote{Note also that we never use the ``volume weighted'' variant of \Eq{eq:FokkerPlanck} proposed as an alternative in \Ref{Linde:1993xx} since then the resulting distribution is not normalizable: its integral over $\dd\phi$ is time- or $N$-dependent. Thus, it leads to probability non-conservation.  Neither is it justified from the physical point of view, since it is based on the assumption that all Hubble physical volumes (``observers'') emerging from the expansion of a previous inflationary patch are clones of each other, while they are strongly correlated.}
\beq
\frac{\partial}{\partial N}P\left(\phi,N\right)=\frac{\partial}{\partial \phi}\left[\frac{V^\prime}{3H^2}P\left(\phi,N\right)\right]+\frac{\partial^2}{\partial\phi^2}\left[\frac{H^2}{8\pi^2}P\left(\phi,N\right)\right]\, .
\label{eq:FokkerPlanck}
\eeq
Now let us compare this equation with the one that would have been obtained if the Langevin equation had been written in terms of cosmic time $t$. Performing the simple change of time variable $\dd N=H\dd t$ in \Eq{eq:Langevin}, this is given by
\beq
\label{eq:Langevin:t}
\frac{\dd\tilde{\varphi}}{\dd t}=-\frac{V^\prime}{3H}+\frac{H^{3/2}}{2\pi}\xi\left(t\right)\, .
\eeq
Here we use the notation $\tilde{\varphi}$ to stress the fact that, a priori, $\tilde{\varphi}$ does not describe the same stochastic process as $\varphi$. The Fokker-Planck equation corresponding to \Eq{eq:Langevin:t} is given by
\beq
\frac{\partial}{\partial t}\tilde{P}\left(\phi,N\right)=\frac{\partial}{\partial \phi}\left[\frac{V^\prime}{3H}\tilde{P}\left(\phi,N\right)\right]+\frac{\partial^2}{\partial\phi^2}\left[\frac{H^3}{8\pi^2}\tilde{P}\left(\phi,N\right)\right]\, .
\label{eq:FokkerPlanck:t}
\eeq
If $H$ is taken to be a function of time only, independent of $\varphi$, the $H$ factors can be taken out of the derivatives with respect to $\phi$ in \Eqs{eq:FokkerPlanck} and~(\ref{eq:FokkerPlanck:t}). In this case, it is straightforward to see that these two are perfectly equivalent through the change of time variable $\dd N=H\dd t$, and that they describe the same stochastic process. On the contrary, if $H$ explicitly depends on $\varphi$, this is obviously no longer the case and $P\neq\tilde{P}$.

This can be better illustrated by calculating the stationary distributions associated with these processes. Let $P_{\mathrm{stat}}(\phi)$ denote a stationary probability distribution for the stochastic process~(\ref{eq:Langevin}), or equivalently,~(\ref{eq:FokkerPlanck}). By definition, $\partial P_{\mathrm{stat}}(\phi)/\partial N=0$, hence
\beq
\frac{\partial}{\partial\phi}\left\lbrace\frac{V^\prime}{3H^2}P_{\mathrm{stat}}\left(\phi\right)+\frac{\partial}{\partial\phi}\left[\frac{H^2}{8\pi^2}P_{\mathrm{stat}}\left(\phi\right)\right]\right\rbrace
\equiv\frac{\partial J}{\partial\phi}=0\, ,
\label{eq:def:J}
\eeq
which defines the probability current $J$. This current thus needs to be independent of $\phi$ for a stationary distribution. In most interesting situations, it is actually $0$. This is notably the case when the allowed values for $\phi$ are unbounded. For example, if $V(\phi)$ is defined up to $\phi=\infty$, the normalization condition $\int P_{\mathrm{stat}}(\phi)\dd\phi=1$ requires that $P_{\mathrm{stat}}(\phi)$ decreases at infinity strictly faster than $\vert\phi\vert^{-1}$. In this case, both $P_{\mathrm{stat}}(\phi)$ and $\partial P_{\mathrm{stat}}(\phi)/\partial\phi$ vanish at infinity. From \Eq{eq:def:J}, $J$ vanishes at infinity also, hence everywhere. This yields a simple differential equation to solve for $P_{\mathrm{stat}}(\phi)$, and one obtains
\beq
\label{eq:stocha:steadystate:N}
P_{\mathrm{stat}}\left(\phi\right)\propto\frac{1}{v\left(\phi\right)}\exp\left[\frac{1}{v\left(\phi\right)}\right]\, .
\eeq
Here, an overall integration constant, which makes the distribution normalized, $\int P_{\mathrm{stat}}(\phi)\dd\phi=1$, is omitted. Similarly, \Eq{eq:FokkerPlanck:t} can be written as $\partial\tilde{P}/\partial t=\partial\tilde{J}/\partial \phi$, and requiring that the current $\tilde{J}$ vanishes gives rise to a differential equation for the stationary distribution $\tilde{P}_\mathrm{stat}(\phi)$, which can easily be solved. One obtains
\beq
\tilde{P}_{\mathrm{stat}}\left(\phi\right)\propto\left[\frac{1}{v\left(\phi\right)}\right]^{3/2}\exp\left[\frac{1}{v\left(\phi\right)}\right]\, .
\label{eq:stocha:steadystate:t}
\eeq
The two distributions are close, and the effects coming from the $H(\varphi)$ dependence are small, only in the regions of the potential where $v\ll  1$.

At this point, we are left with the issue of identifying the right time variable to work with. Actually, one can explicitly show~\cite{Finelli:2008zg, Finelli:2010sh, Finelli:2011gd} that $N$ is the correct answer, and that it is the only time variable that allows the stochastic formalism to reproduce a number of results from QFT on curved space-times. We leave this discussion to appendix~\ref{sec:app:whyN}, where we elaborate on existing results and show why, since we deal with metric perturbations, we must work with $N$.
\section{Method}
\label{sec:PSinstocha}
Let us now review how correlation functions of curvature fluctuations can be calculated in stochastic inflation, and see which approach is best suited to the issue we are interested in.

The problem can first be treated at linear order \cite{Matacz:1996fv, Liguori:2004fa, Kunze:2006tu} by expanding the coarse grained field $\varphi$ about its classical counterpart $\phi_\ucl$, $\varphi= \phi_\ucl+ \delta\phi^{(1)}$. Here, recall that $\phi_\ucl$ is the solution of the Langevin equation~(\ref{eq:Langevin}) without the noise term. The quadratic moment of $\delta\phi^{(1)}$ can be calculated as in appendix~\ref{sec:sto:StoAndQFT}, see \Eq{eq:stocha:deltaphi1Squared}. It corresponds to the integrated power spectrum of the field fluctuations on super-Hubble scales, and can therefore be related~\cite{Kunze:2006tu} to the power spectrum $\calP_\zeta$ of curvature perturbations thanks to the relation
\beq
\calP_\zeta\simeq \frac{\dd}{\dd N}\left\lbrace\left(\frac{\dd\phi_\ucl}{\dd N}\right)^{-2}\left\langle\left[\delta\phi^{(1)}\right]^2 \right\rangle\right\rbrace\, .
\label{eq:stocha:powerspectrum:pert}
\eeq
In this expression, the right hand side needs to be evaluated when the scale associated with the wavenumber $k$ (at which the power spectrum is calculated) exits the Hubble radius. If one plugs the expression~(\ref{eq:stocha:deltaphi1Squared}) obtained in appendix~\ref{sec:app:whyN} for $\langle\delta{\phi^{(1)}}^2 \rangle$ using $N$ as the time variable into \Eq{eq:stocha:powerspectrum:pert}, one obtains
\beq
\calP_\zeta\simeq \frac{H^2(\phi_\ucl)}{8\pi^2\Mp^2\epsilon_1\left(\phi_\ucl\right)}\, .
\label{eq:PS:pertstocha}
\eeq
As before, $\phi_\ucl$ needs to be evaluated when the scale associated with the wavenumber at which the power spectrum is calculated exits the Hubble radius. The quantity $\epsilon_1\equiv-\dd{H}/(H^2\dd t)$ is the first slow-roll parameter. At leading order in slow roll, it verifies $\epsilon_1=\Mp^2/2(v^\prime/v)^2$. The above expression exactly matches the standard result~\cite{Mukhanov:1985rz, Mukhanov:1988jd}. In order to get the first corrections to this standard result, one thus needs to go to higher orders in $\delta\phi$. Actually, one can show that no contributions arise at next-to-leading order, and that one needs to go at least to next-to-next-to-leading order. This renders the calculation technically difficult. This is why we will prefer to make use of non-perturbative techniques. 
In passing, let us stress that in \Ref{Kunze:2006tu}, the Langevin equation is written and  solved with $t$, whereas, as already said, the number of \efolds $N$ must be used instead. This has important consequences. Indeed, if one makes use of cosmic time $t$ and plugs the associated expression~(\ref{eq:stocha:deltaphi1Squared:t}) for the quadratic moment of $\delta\tilde{\phi}^{(1)}$ into \Eq{eq:stocha:powerspectrum:pert}, one obtains
\beq
{\calP}_{\tilde{\zeta}} \simeq \frac{H^2(\phi_\ucl)}{8\pi^2\Mp^2\epsilon_1\left(\phi_\ucl\right)}\left\lbrace 1+2\left[\frac{H^\prime\left(\phi_\ucl\right)}{H\left(\phi_\ucl\right)}\right]^2 \int^{\phi_\ucl} \left[\frac{H\left(\phi_\ucl\right)}{H^\prime\left(\phi_\ucl\right)}\right]^3\dd\phi\right\rbrace\, .
\label{eq:stocha:powerspectrum:perturbative:t}
\eeq
Here, we have adopted the same notation as in section~\ref{sec:statDistrib} where a tilde recalls that not the same quantity is worked out and $\tilde{\zeta}$ is not $\zeta$. This result matches Eq.~(2.11) of \Ref{Kunze:2006tu}. However, in this work, it is concluded that, because of the second term in the braces of \Eq{eq:stocha:powerspectrum:perturbative:t}, which is always negative, the amplitude of the power spectrum in the stochastic approach is in general reduced with respect to the standard result. One can see that such a statement is incorrect, since the extra term in \Eq{eq:stocha:powerspectrum:perturbative:t} is simply due to not working with the correct time variable. This is why, if such an approach were to be followed and extended to higher orders, it would again be crucial to work with $N$ as the time variable.

Another strategy is followed in \Refs{Kuhnel:2008yk, Kuhnel:2008wr, Kuhnel:2010pp}, where methods of statistical physics, such as replica field theory, are employed in a stochastic inflationary context. However, only the case of a free test field evolving in a de Sitter or power-law background is investigated, while we need to go beyond the fixed background assumption in order to study the effects of the explicit $H(\varphi)$ dependence. This is why we cannot directly make use of this computational scheme in the present work.

Finally, in \Refs{Enqvist:2008kt, Fujita:2013cna, Fujita:2014tja}, the $\delta N$ formalism is used to relate the curvature perturbations to the number of \efolds statistics. This is this last route that we chose to follow here, since it does not rely on any perturbative expansion scheme, and since it does not prevent us from implementing the explicit $H(\varphi)$ dependence. In \Ref{Fujita:2014tja}, numerical solutions are obtained for quadratic and hybrid potentials. In the present work, we derive fully analytical and non-perturbative results that apply to any single-field potential, and which do not require a numerical solution of the Langevin equation. As a by-product, this allows us to prove, for the first time, that the standard results are always recovered in the classical limit, for any potential. 
\subsection[The $\delta N$ Formalism]{The \texorpdfstring{$\bm{\delta N}$}{$\delta N$} Formalism}
\label{sec:deltaN}
The $\delta N$ formalism~\cite{Starobinsky:1982ee, Starobinsky:1986fxa, Sasaki:1995aw, Sasaki:1998ug, Lyth:2004gb, Lyth:2005fi} is very well suited to addressing the calculation of correlation functions in stochastic inflation, since it relates the statistical properties of curvature perturbations to the distribution of the number of \efolds among a family of homogeneous universes. Let us first recall where this correspondence comes from and, as an example, how the scalar power spectrum is usually calculated in the associated formalism. 

Starting from the unperturbed flat Friedmann-Lema\^itre-Robertson-Walker (FLRW) line element, $\dd s^2=-\dd t^2+a^2(t)\delta_{ij}\dd x^i \dd x^j$, deviations from homogeneity and isotropy can be included in a more general, perturbed metric, which contains some gauge redundancy. A specific gauge choice consists in requiring that fixed $t$ slices of space-time have uniform energy density, and that fixed $x$ worldlines be comoving. When doing so, and including scalar perturbations only, the perturbed metric in this gauge (which coincides in the super-Hubble regime with the synchronous gauge supplemented by some additional conditions fixing it uniquely) becomes~\cite{Starobinsky:1982ee, Creminelli:2004yq,Salopek:1990jq} $\dd s^2=-\dd t^2+ a^2(t) \ee^{2\zeta(\bm{x})}\delta_{ij}\dd x^i \dd x^j$, up to small terms proportional to gradients of $\zeta$. Here, $\zeta$ is the adiabatic (curvature) perturbation, which is time-independent in single-field inflation once the decaying mode can be neglected. The omission of tensor perturbations is justified by the fact that their amplitude is suppressed compared to the scalar ones by the small slow-roll parameter $\epsilon_1$. This allows us to define a local scale factor $\tilde{a}(t,\bm{x})=a(t)\ee^{\zeta(\bm{x})}$. Starting from an initial flat slice of space-time at time $t_\uin$, the amount of expansion $N(t,\bm{x})\equiv\ln\left[\tilde{a}(t,\bm{x})/a(t_\uin)\right]$ to a final slice of uniform energy density is then related to the curvature perturbation through
\beq
\label{eq:zeta:deltaN}
\zeta(\bm{x})=N\left(t,\bm{x}\right)-N_0(t)\equiv \delta N\, ,
\eeq
where $N_0(t)\equiv\ln\left[a(t)/a(t_\uin)\right]$ is the unperturbed amount of expansion. From this, an important simplification arises on large scales where anisotropy and spatial gradients can be neglected, and the local density, expansion rate, \etc , obey the same evolution equations as a homogeneous FLRW universe. Thus we can use the homogeneous FLRW solutions to describe the local evolution, which is known as the ``quasi-isotropic''~\cite{Lifshitz:1960,Starobinsky:1982mr,Comer:1994np,Khalatnikov:2002kn} or ``separate universe'' approach~\cite{Wands:2000dp,Lyth:2003im,Lyth:2004gb}. It implies that $N(t,\bm{x})$ is the amount of expansion in unperturbed, homogeneous universes, so that $\zeta$ can be calculated from the knowledge of the evolution of a family of such universes. Written in terms of the inflaton field $\phi(\bm{x})=\phi+\delta\phi(\bm{x})$, consisting of an unperturbed, homogeneous piece $\phi$ and of a perturbation $\delta \phi$ originating from quantum fluctuations, \Eq{eq:zeta:deltaN} gives rise to
\beq
\label{eq:zeta:deltaN:2}
\zeta\left(\bm{x}\right)=N\left[\rho\left(t\right),\phi\left(\bm{x}\right)\right]-N\left[\rho\left(t\right),\phi\right]\, .
\eeq
Here, $N$ is to be evaluated in unperturbed universes from an initial epoch when the inflaton field has an assigned value $\phi$ to a final epoch when the energy density has an assigned value $\rho$. Since the observed curvature perturbations are almost Gaussian, at leading order in perturbation theory, one has
\beq
\label{eq:deltaNform:DeltaNDeltaPhi}
\zeta\left(\bm{x}\right)=\delta N\simeq \frac{\partial N}{\partial \phi}\delta\phi\, .
\eeq
Here, $N\left(\phi\right)$ is usually evaluated with the slow-roll, classical formula
\beq
\label{eq:deltaNform:traj:class}
N\left(\phi\right)=\frac{1}{\Mp}\int\frac{\dd\phi}{\sqrt{2 \epsilon_1}}\, .
\eeq
Once $\zeta$ is decomposed into Fourier components, $\zeta_{\bm{k}}=(2\pi)^{-3/2}\int\dd^3\bm{x}\zeta(\bm{x})\exp(i\bm{k}\cdot\bm{x})$, the power spectrum $\mathcal{P}_\zeta$ is defined from the quantum expectation value $\mean{\zeta_{\bm{k}}\zeta_{{\bm{k}^\prime}}}\equiv (2\pi)^5 / (2k^3)$ $\mathcal{P}_\zeta(k) \delta( \bm{k} + {\bm{k}^\prime})$. It can be expressed in terms of the power spectrum of $\delta\phi$ (defined by similar relations) thanks to \Eq{eq:deltaNform:DeltaNDeltaPhi}. For quasi-de Sitter inflation, and when the curvature of the inflaton potential is much smaller than $H$, on super-Hubble scales, the latter is given by~\cite{Bunch:1978yq} $\mathcal{P}_{\delta\phi}(k)\simeq H^2(k)/4\pi^2$, where $H(k)$ means $H$ evaluated at the time when the $k$ mode crosses the Hubble radius, \ie when $aH=k$. Together with \Eq{eq:deltaNform:traj:class}, one therefore obtains
\beq
\mathcal{P}_{\zeta}=\left[\frac{H(k)}{2\pi}\right]^2\frac{1}{2\Mp^2\epsilon_1\left(k\right)}\, ,
\label{eq:deltaNform:Pzeta}
\eeq
which is the same as \Eq{eq:PS:pertstocha} and which matches the standard result~\cite{Mukhanov:1985rz, Mukhanov:1988jd}.

A fundamental remark is that in the above usual calculation, the quasi-isotropic (separate universe) approximation is assorted with the assumption that on super-Hubble scales, the evolution of the inflaton field is governed by its classical equation of motion~(\ref{eq:deltaNform:traj:class}). The stochastic dispersion in the number of \efolds thus only comes from the field dispersion at Hubble crossing $\delta\phi_*$. In most cases, this is a good approximation for the following reason. From the Langevin equation~(\ref{eq:Langevin}), one can see that during the typical time scale of one \efold, the classical drift of the inflaton field is of the order $\Delta\phi_\ucl=V^\prime/(3H^2)=\sqrt{2\epsilon_1}\Mp$, while the quantum kick is of the order $\Delta\phi_{\mathrm{qu}}=H/(2\pi)$. This allows us to define a rough ``classicality'' criterion $\Delta\phi_{\mathrm{qu}}/\Delta\phi_\ucl$ that assesses the amplitude of the stochastic corrections to the classical trajectory. Making use of \Eqs{eq:deltaNform:Pzeta}, this ratio can be expressed as
\beq
\frac{\Delta\phi_{\mathrm{qu}}}{\Delta\phi_\ucl}=\sqrt{\mathcal{P}_\zeta}\, ,
\label{eq:stocha:eta:Pzeta}
\eeq
which is valid for single-field slow-roll models of inflation with canonical kinetic terms. Since $\mathcal{P}\sim 2\times 10^{-9}$ for the modes observed in the Cosmic Microwave Background (CMB), stochastic effects are already small when these modes cross the Hubble radius. If one further assumes that $\epsilon_1$ monotonously grows toward $1$ during the last stages of inflation, $\mathcal{P}_\zeta\propto H^2/\epsilon_1$ decreases (since $H$ can only decrease) and one is therefore ensured that the stochastic corrections to the inflaton trajectory remain small.

However, they are two caveats to this line of reasoning. The first one is that, as we will show below, $\Delta\phi_{\mathrm{qu}}/\Delta\phi_\ucl$ is not the correct way to assess the importance of stochastic effects and one should use instead another classicality criterion that we will derive. The second one is that, in some situations, $\epsilon_1$ becomes tiny or even vanishes in some transient phase between the Hubble exit time of the observed modes and the end of inflation. This is the case, for example, when the potential has a flat inflection point, such as in MSSM inflation~\cite{Lyth:2006ec, Allahverdi:2008bt, Enqvist:2010vd} or as in punctuated inflation~\cite{Jain:2008dw, Jain:2009pm}. Another situation of interest is when inflation does not have a graceful exit but ends due to tachyonic instability involving an auxiliary field, like in hybrid inflation~\cite{Linde:1993cn,Martin:2011ib}, or by brane annihilation in string-theoretical setups~\cite{Dvali:1998pa, Alexander:2001ks}. In such cases, $\epsilon_1$ can decrease and the last \efolds of inflation may be dominated by the quantum noise. It is therefore important to study the dispersion $\delta N$ arising not only from $\delta\phi_*$ but from the complete subsequent stochastic history of the coarse grained field.

Note also that in these expressions, $\zeta$ need not be small as was shown in \Refs{Starobinsky:1982ee,Lyth:2004gb,Naruko:2011zk} [note, however, that $\zeta$ is defined up to a constant due to an arbitrary possible rescaling of $a(t)$], thus, $\delta N$ need not be small, too. As follows from the quasi-isotropic (separate
universe) approach, the condition for inflation to proceed is only that $H\ll \Mp$. On the other hand, if $P_{\zeta}(k)\sim H/(\Mp \sqrt{\epsilon_1})$ exceeds unity (the so called regime of ``eternal inflation''), then the Universe loses its local homogeneity and isotropy after the end of inflation, but not immediately. This occurs much later than the comoving scale $a(t)/k$ at which this inhomogeneity occurs crosses the Hubble radius $H^{-1}$ second time. Thus, in the scope of the inflationary scenario $P_{\zeta}$ may well exceed unity at scales much exceeding the present Hubble radius. The stochastic inflation approach provides us with a possibility to obtain quantitatively correct results in this non-linear regime, too. 
\subsection{Computational Programme}
\label{sec:programme}
This is why we now generalize this approach to a fully stochastic framework. For a given wavenumber $k$, let $\phi_*(k)$ be the mean value of the coarse grained field when $k$ crosses the Hubble radius. If inflation terminates at $\phi_\uend$, let $\mathcal{N}(k)$ denote the number of \efolds realized between $\phi_*(k)$ and $\phi_\uend$. Obviously, $\mathcal{N}$ is a stochastic quantity, and we can define its variance
\beq
\delta\mathcal{N}^2\left(k\right)\equiv\left\langle \mathcal{N}^2\left(k\right)\right\rangle - \left\langle \mathcal{N}\left(k\right)\right\rangle^2\, .
\label{eq:deltacalN:def}
\eeq
It is related with the curvature perturbation $\delta N$ of \Eq{eq:deltaNform:DeltaNDeltaPhi} in the following manner. Since $\delta\mathcal{N}$ is computed between two fixed points $\phi_*\left(k\right)$ and $\phi_\uend$, it receives an integrated contribution from all the modes crossing the Hubble radius between these two points, and one has
\beq
\delta\mathcal{N}^2\left(k\right)=\int_{k}^{k_\uend}\mathcal{P}_{\delta N}\left(k\right) \dfrac{\dd k}{k}=\displaystyle\int_{\ln k_\uend-\left\langle\mathcal{N}\right\rangle\left(1-\epsilon_{1*}+\cdots\right)}^{\ln k_\uend}\mathcal{P}_{\delta N}\dd N\, .
\eeq
Here we have used the relation $\langle \mathcal{N}(k)\rangle =\ln(a_\uend/a_*)=\ln(k_\uend/k) (1+\epsilon_{1*}+\cdots)$, where $\epsilon_{1*}+\cdots$ stand for slow-roll corrections that we do not need to take into account at leading order in slow roll. One then has
\beq
\label{eq:stocha:PS}
\mathcal{P}_\zeta\left(k\right)=\mathcal{P}_{\delta\mathcal{N}}\left(k\right)=\left.\dfrac{\dd  \delta\mathcal{N}^2}{\dd  \left\langle \mathcal{N} \right\rangle}\right\vert_{\left\langle\mathcal{N}\right\rangle=\ln\left(k_\uend/k\right)}\, .
\eeq
In the same manner, the third moment of the number of \efolds distribution,
\beq
\delta\mathcal{N}^3\left(k\right)\equiv \langle\left(\mathcal{N}-\langle\mathcal{N}\rangle\right)^3\rangle\, ,
\label{eq:deltaN3:def}
\eeq
receives a double integrated contribution from the local bispectrum $\mathcal{B}_\zeta$, and one has $\mathcal{B}_\zeta\propto \dd^2\delta\mathcal{N}^3/\dd\langle \mathcal{N}\rangle^2$. The local $\fnl$ parameter, measuring the ratio between the bispectrum and the power spectrum squared, is then given by
\beq
\fnl=\frac{5}{72}\frac{\dd^2 \delta\mathcal{N}^3}{\dd\langle\mathcal{N}\rangle^2}\left(\frac{\dd \delta\mathcal{N}^2}{\dd\langle\mathcal{N}\rangle} \right)^{-2}\, ,
\label{eq:stocha:fnl}
\eeq
where $5/72$ is a conventional historical factor. Analogously, the trispectrum is related to the third derivative of $\delta\mathcal{N}^4$ with respect to $\langle\mathcal{N}\rangle$, and so on and so forth.

The computational programme we must follow is now clear. For a given mode $k$, we first calculate $\phi_*(k)$ (this sets the location of the observational window). We then consider stochastic realizations of \Eq{eq:Langevin} that satisfy $\varphi=\phi_*(k)$ at some initial time,\footnote{This calculation therefore relies on a specific choice of initial (in fact, pre-inflationary) conditions, since all trajectories emerge from $\phi_*$ at initial time. In principle, other choices could be made, even if most physical quantities (in particular, perturbations during the last $60$ $e$-folds) do not depend on them.} and denote by $\mathcal{N}$ the number of \efolds that is realized before reaching $\phi_\uend$. Among these realizations, we calculate the first moments of this stochastic quantity, $\langle \mathcal{N} \rangle$, $\langle \mathcal{N}^2 \rangle$, $\langle \mathcal{N}^3 \rangle$, {\etc} We finally apply relations such as \Eqs{eq:stocha:PS} and~(\ref{eq:stocha:fnl}) to obtain the power spectrum, the non-Gaussianity local parameter, or any higher order correlation function. 
\subsection{First Passage Time Analysis}\label{sec:FirstPassageTimeAnalysis}
\begin{figure}[t]
\begin{center}
\includegraphics[width=\wdblefig]{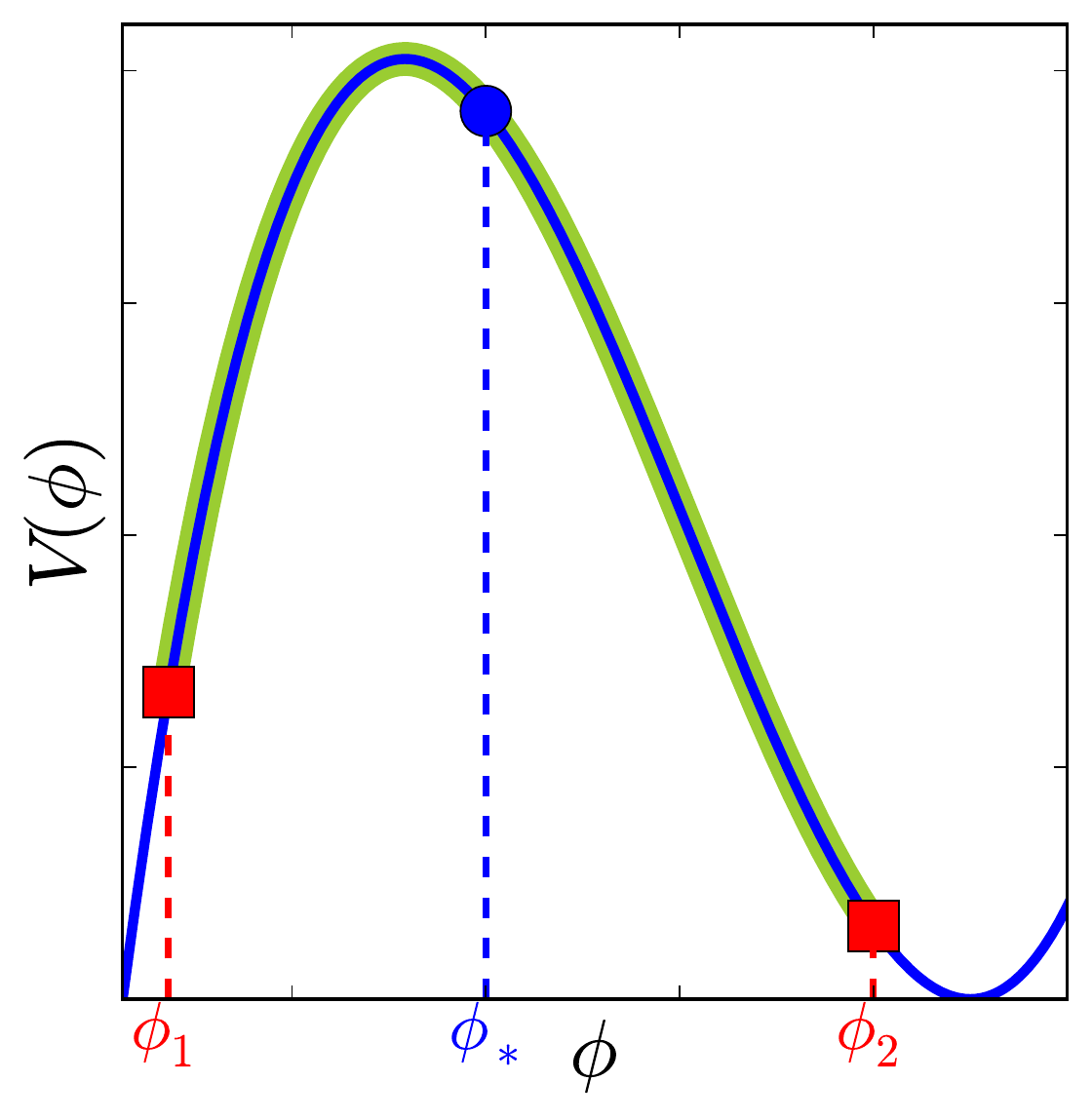}
\includegraphics[width=\wdblefig]{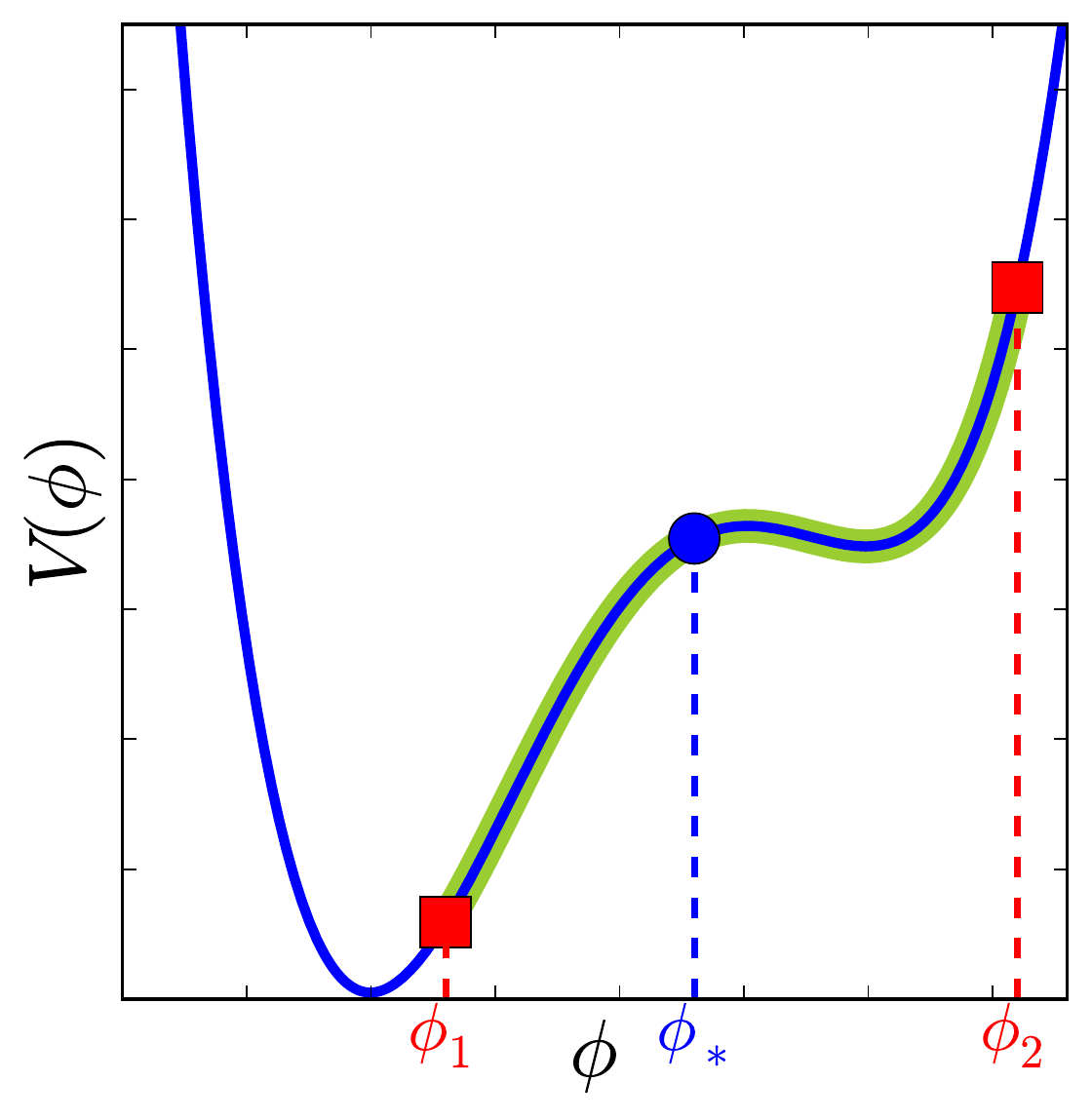}
\caption[Sketch of the Langevin bounded dynamics]{Sketch of the dynamics solved in section~\ref{sec:PSinstocha}. The inflaton is initially located at $\phi_*$ and evolves along the potential $V(\phi)$ under the stochastic Langevin equation~(\ref{eq:Langevin}), until it reaches one of the two ending  values $\phi_1$ or $\phi_2$. The left panel is an example where inflation always terminates by slow-roll violation, while the right panel stands for a situation where one of the ending points, $\phi_2$, corresponds to where $V\sim \Mp^4$ above which inhomogeneities prevent inflation from occurring.}
\label{fig:sketch}
\end{center}
\end{figure}

In what follows, this calculation is performed using the techniques developed in ``first passage time analysis''~\cite{Bachelier:1900, Gihman:1972}, which was applied to stochastic inflation in \Ref{Starobinsky:1986fx}. We consider the situation sketched in \Fig{fig:sketch}, where the inflaton is initially located at $\phi_*$ and evolves in some potential $V\left(\phi\right)$ under \Eq{eq:Langevin}. 
Because any part of the potential can a priori be explored, here we consider two possible ending points, $\phi_1$ and $\phi_2$, located on each side of $\phi_*$. If the potential is, say, of the hilltop type (left panel), $\phi_1$ and $\phi_2$ can be taken at the two values where inflation has a graceful exit, on each side of the maximum of the potential. If, on the other hand, a flat potential extends up to $\phi=\infty$ (right panel), one of these points, say $\phi_2$, can be taken where $V$ becomes super-Planckian and inhomogeneities prevent inflation from occurring. In such cases, the precise value of $\phi_2$ plays a negligible role, as we will show in section~\ref{sec:WallHittProba}. Let $\mathcal{N}$ be the number of \efolds realized during this process.

Before proceeding with the calculation of the $\mathcal{N}$ moments, a first useful result to establish is the It\^o lemma, which is a relation verified by any smooth function $f$ of $\varphi$. The Taylor expansion of such a function at second order is given by $f\left(\varphi+\dd \varphi\right)=f\left(\varphi\right)+f^{\prime}\left(\varphi\right)\dd \varphi+f^{\prime\prime}\left(\varphi\right)/2\,\dd \varphi^2+\mathcal{O}\left(\dd \varphi^3\right)$. Now, if $\varphi$ is a realization of the stochastic process under study, $\dd \varphi$ is given by \Eq{eq:Langevin} and at first order in $\dd N$, one obtains
\bea
\dd f\left[\varphi\left(N\right)\right]&=&
f^{\prime}\left[\varphi\left(N\right)\right]{\sqrt{2v\left[\varphi\left(N\right)\right]}}\Mp\xi\left(N\right)\dd N
\nonumber\\& &
-f^{\prime}\left[\varphi\left(N\right)\right]\frac{v^\prime\left[\varphi\left(N\right)\right]}{v\left[\varphi\left(N\right)\right]}\Mp^2\dd N +\Mp^2 f^{\prime\prime}\left[\varphi\left(N\right)\right]v\left[\varphi\left(N\right)\right]\dd N\, .
\eea
Integrating this relation between $N=0$ where $\varphi=\phi_*$ and $N=\mathcal{N}$ where $\varphi=\phi_1$ or $\phi_2$, one gets the It\^o lemma~\cite{ito1944}
\bea
&&f\left(\phi_1\ \mathrm{or}\ \phi_2\right)-f\left(\phi_*\right)=
\int_0^{\mathcal{N}} f^{\prime}\left[\varphi\left(N\right)\right]\sqrt{2v\left[\varphi\left(N\right)\right]}\Mp\xi\left(N\right)\dd N
\nonumber\\&&\ \
+\int_0^{\mathcal{N}} \left\lbrace \Mp^2 f^{\prime\prime}\left[\varphi\left(N\right)\right]v\left[\varphi\left(N\right)\right]
-f^{\prime}\left[\phi\left(N\right)\right]\frac{v^\prime\left[\varphi\left(N\right)\right]}{v\left[\varphi\left(N\right)\right]}\Mp^2
\right\rbrace\dd N\, ,
\nonumber\\
\label{eq:ito}
\eea
which we will repeatedly make use of in the following.
\subsubsection{Ending Point Probability}\label{sec:WallHittProba}
As a first warm-up, let us calculate the probability $p_1$ that the inflaton field first reaches the ending point located at $\phi_1$ [\ie $\phi\left(\mathcal{N}\right)=\phi_1$], or, equivalently the probability $p_2=1-p_1$ that the inflaton field first reaches the ending point located at $\phi_2$ [\ie $\phi\left(\mathcal{N}\right)=\phi_2$]. This will also allow us to determine when the ending point located at $\phi_2$ plays a negligible role.

First of all, let $\psi\left(\varphi\right)$ be a function of the coarse grained field that can be expressed as
\beq
\psi\left(\varphi\right)=\frac{h\left(\varphi\right)-h\left(\phi_2\right)}{h\left(\phi_1\right)-h\left(\phi_2\right)}\, ,
\label{eq:psi:h}
\eeq
where $h\left(\varphi\right)$ will be specified later. By construction, one has $\psi\left(\phi_1\right)=1$ and $\psi\left(\phi_2\right)=0$. This implies that the mean value of $\psi$ evaluated at $\varphi\left(\mathcal{N}\right)$ is given by $\left\langle \psi\left[\varphi\left(\mathcal{N}\right)\right] \right\rangle= p_1\psi\left(\phi_1\right) + p_2\psi\left(\phi_2\right)=p_1$. The idea is then to find an $h\left(\varphi\right)$ function that makes easy the evaluation of the left hand side of the previous relation, so that we can deduce $p_1$. In order to do so, let us apply the It\^o lemma~(\ref{eq:ito}) to $h\left(\varphi\right)$. If one requires that the integral of the second line of \Eq{eq:ito} vanishes, that is,
\beq
h^{\prime\prime}\left(\varphi\right)v\left(\varphi\right)=h^\prime\left(\varphi\right)\frac{v^\prime\left(\varphi\right)}{v\left(\varphi\right)}\, ,
\label{eq:hittwall:h:equadiff}
\eeq
one obtains
\beq
h\left[\varphi\left(\mathcal{N}\right)\right]-h\left(\phi_\uin\right)=
\int_0^{\mathcal{N}} h^{\prime}\left[\varphi\left(N\right)\right]\sqrt{2v\left[\varphi\left(N\right)\right]}\Mp\xi\left(N\right)\dd N\, .
\label{eq:h:ito}
\eeq
Because $h$ and $\psi$ are linearly related, see \Eq{eq:psi:h}, the same equation is satisfied by $\psi$. When averaged over all realizations,\footnote{The fact that the averaged integral in the right hand side of \Eq{eq:h:ito} vanishes is non-trivial since both the integrand and the upper bound are stochastic quantities, but this can be shown rigorously (see \eg p. 12 of \Ref{Gihman:1972}).} its right hand side vanishes. One then obtains $\left\langle \psi\left[\varphi\left(\mathcal{N}\right)\right] \right\rangle=\psi\left(\phi_*\right)$, which is the probability $p_1$ one is seeking for. All one needs to do is therefore to solve \Eq{eq:hittwall:h:equadiff} to obtain $h\left(\varphi\right)$, to plug the obtained expression into \Eq{eq:psi:h} to derive $\psi(\varphi)$, and finally to evaluate this function at $\phi_*$. A formal solution to \Eq{eq:hittwall:h:equadiff} is given by $h\left(\varphi\right)=A\int_{B}^{\varphi}\exp\left[-1/v\left(x\right)\right]\dd x$, where $A$ and $B$ are two integration constants that play no role, since they cancel out when calculating $\psi$ thanks to \Eq{eq:psi:h}. Indeed, the latter gives rise to
\beq
p_1
=\dfrac{\displaystyle\int_{\phi_*}^{\phi_2}\exp\left[-\frac{1}{v\left(x\right)}\right]\dd x}{\displaystyle\int_{\phi_1}^{\phi_2}\exp\left[-\frac{1}{v\left(x\right)}\right]\dd x}
\, ,
\label{stocha:hittproba:p1}
\eeq
and a symmetric expression for $p_2$.\footnote{This is in agreement with Eq. (29) of \Ref{Starobinsky:1986fx}, derived in the case where $H$ is constant, hence $v^{-1}\approx v_*^{-1} - (v-v_*)\,v_*^{-2}$, where $\phi_2$ and $\phi_1$ lie at $\pm\infty$ correspondingly, and where the initial condition for \Eq{eq:FokkerPlanck} is chosen to be $P(\phi,0)=\delta(\phi - \phi_*)$.}

A few remarks are in order about this result. First, one can check that, since $\phi_*$ lies between $\phi_1$ and $\phi_2$, the probability~(\ref{stocha:hittproba:p1}) is ensured to be between $0$ and $1$. Second, one can also verify that when $\phi_*=\phi_1$, $p_1=1$, and when $\phi_*=\phi_2$, $p_1=0$, as one would expect. Third, in the case depicted in the right panel of \Fig{fig:sketch}, in the limit where $\phi_2\rightarrow\infty$, one is sure to first reach the ending point located at $\phi_1$, that is, $p_2=\int_{\phi_1}^{\phi_*}e^{-1/v}/\int_{\phi_1}^{\phi_2}e^{-1/v}=0$. Indeed, the numerator of the expression for $p_2$ is finite, since a bounded function is integrated over a bounded interval. If the potential is maximal at $\phi_2$, and if it is monotonous over an interval of the type $\left[\phi_0,\phi_2\right[$, its denominator is on the contrary larger than the integral of a function bounded from below by a strictly positive number, over an unbounded interval $\left[\phi_0,\phi_\infty\right[$. This is why it diverges, and why $p_2$ vanishes. This means that if $\phi_2$ is sufficiently large, its precise value plays no role, since inflation always terminates at $\phi_1=\phi_\uend$.
\subsubsection[Mean Number of \efolds]{Mean Number of \texorpdfstring{$\bm{e}$}{$e$}-folds}\label{sec:meanN}
\begin{figure}[t]
\begin{center}
\includegraphics[width=\wdblefig]{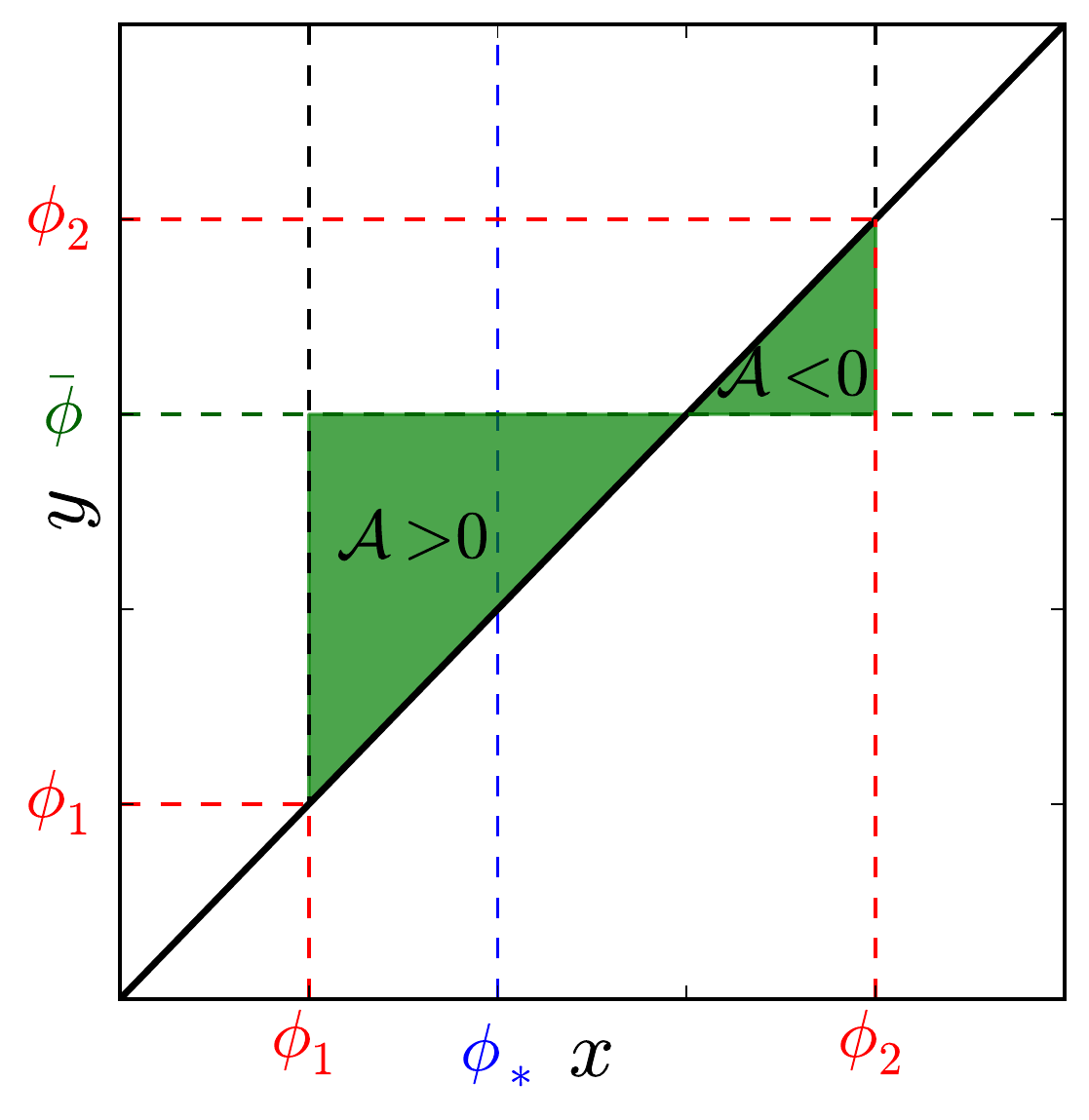}
\caption[Integration domain of the mean number of \efolds]{Integration domain of \Eq{eq:f:sol} when evaluated at $\varphi=\phi_2$, in the case $\phi_1<\phi_2$ (the opposite case proceeds the same way). The discrete parameter $x$ is integrated between $\phi_1$ and $\phi_2$, while $y$ varies between $x$ and $\bar{\phi}$. The resulting integration domain is displayed in green. When $x<\bar{\phi}$, one has $\dd x\dd y >0$ and one integrates a positive contribution to the mean number of $e$-folds. Conversely, when $x>\bar{\phi}$, one has $\dd x\dd y <0$ and one integrates a negative contribution. This is necessary in order for the overall integral to vanish. This is why $\bar{\phi}$ must lie between $\phi_1$ and $\phi_2$.}
\label{fig:NmeanIntegrationDomain}
\end{center}
\end{figure}
\begin{figure}[t]
\begin{center}
\includegraphics[width=0.474\textwidth]{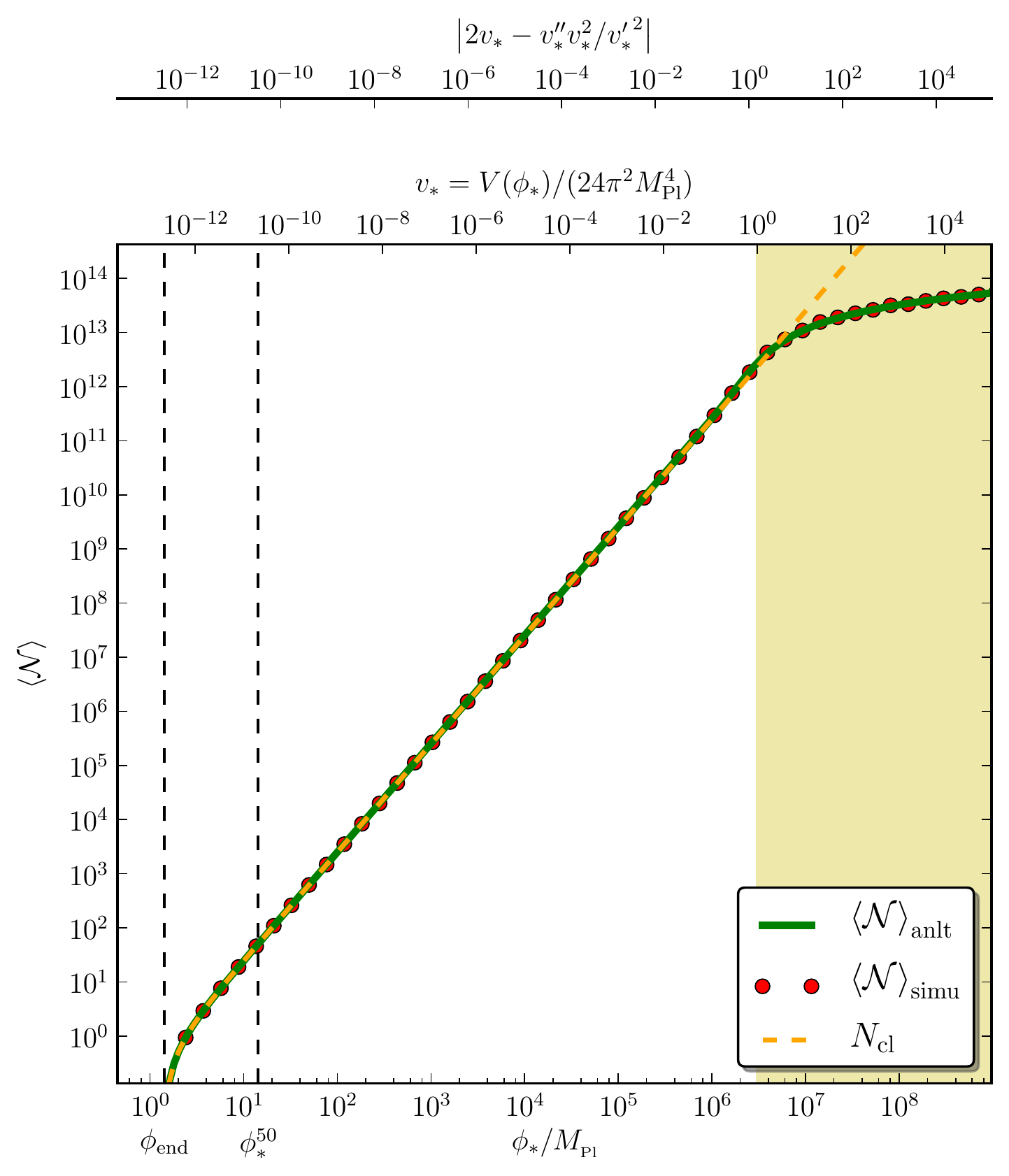}
\includegraphics[width=0.509\textwidth]{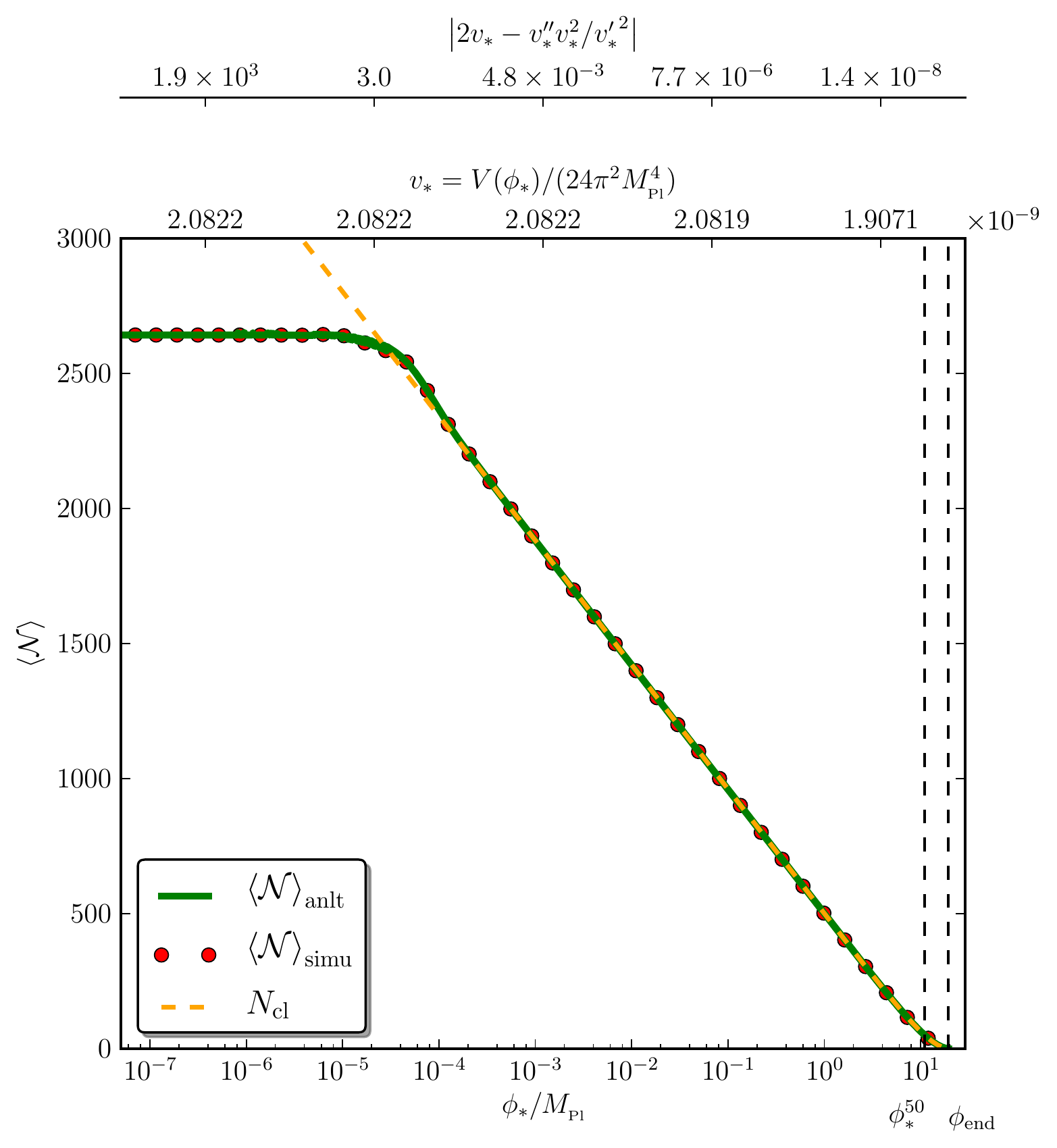}
\caption[mean number of \efolds]{Mean number of \efolds $\langle \mathcal{N}\rangle $ realized in the large field $V\propto\phi^2$ (left panel) and small field $V\propto 1-\phi^2/\mu^2$ (where $\mu=20\Mp$, right panel) potentials, as a function of $\phi_*$. The location $\phi_*^{50}$ refers to the value of $\phi_*$ for which the classical number of \efolds $N_\ucl=50$ and $\phi_\uend$ is where $\epsilon_1=1$. In both panels, the overall mass scale in the potential is set to the value that fits the observed amplitude of the power spectrum $\calP_\zeta\sim 2\times 10^{-9}$, when the latter is calculated with the classical formula~(\ref{eq:deltaNform:Pzeta}), $50$ \efolds before the end of inflation. 
The green line corresponds to the analytical exact result~(\ref{eq:Nmean}), and the red circles are provided by a numerical integration of the Langevin equation~(\ref{eq:Langevin}) for a large number of realizations over which the mean value of $\mathcal{N}$ is computed. The orange dashed line corresponds to the classical limit~(\ref{eq:stocha:meanN:classtraj}). The top axes display $v$ and the classicality criterion $\vert 2v-v^{\prime\prime}v^2/{v^\prime}^2\vert$. The yellow shaded area stands for $v > 1$, where inhomogeneities are expected to prevent inflation from occurring and our calculation cannot be trusted anymore.}
\label{fig:Nmean}
\end{center}
\end{figure}

Let us now turn to the calculation of the mean number of \efolds $\langle\mathcal{N}\rangle$. As above, we want to make use of the It\^o lemma~(\ref{eq:ito}). To do so, let us define $f(\varphi)$ as the solution of the differential equation
\beq
\label{eq:v:mean} f^{\prime\prime}\left(\varphi\right) v\left(\varphi\right)
-f^{\prime}\left(\varphi\right)\frac{v^\prime\left(\varphi\right)}{v\left(\varphi\right)}=-\frac{1}{\Mp^2}\, ,
\eeq
with boundary conditions $f\left(\phi_1\right)=f\left(\phi_2\right)=0$. Such a solution will be explicitly calculated in due time. For now, it is interesting to notice that when this is plugged into  the It\^o lemma~(\ref{eq:ito}), the first term of the left hand side, $f\left(\phi_1\ \mathrm{or}\ \phi_2\right)$, vanishes, and the second integrand of the right hand side is $-1$. Thus, the It\^o equation can be rewritten as
\beq
\label{eq:stild:stocha}
\mathcal{N}=f\left(\phi_*\right)+\int_0^{\mathcal{N}} f^{\prime}\left[\varphi\left(N\right)\right]\sqrt{2v\left[\varphi\left(N\right)\right]}\Mp\xi\left(N\right)\dd N
\, .
\eeq
By averaging over realizations, one obtains\footnote{Here again, since both the integrand and the upper bound are stochastic quantities, it is non-trivial that the integral in the right hand side of \Eq{eq:stild:stocha} vanishes when averaged, but it can be shown rigorously.}
\beq
\label{eq:meantime}
\left\langle\mathcal{N}\right\rangle=f\left(\phi_*\right)\, .
\eeq
What one needs to do is therefore to solve the deterministic differential equation~(\ref{eq:v:mean}) with the associated boundary conditions, and to evaluate the solution at $\phi_*$. One obtains
\beq
f\left(\varphi\right)=\int^{\varphi}_{\phi_1}\frac{\dd x}{\Mp}\int^{\bar{\phi}\left(\phi_1,\phi_2\right)}_{x}\frac{\dd y}{\Mp}\frac{1}{v\left(y\right)}\exp \left[\frac{1}{v\left(y\right)}-\frac{1}{v\left(x\right)}\right]\, ,
\label{eq:f:sol}
\eeq
where $\bar{\phi}$ is an integration constant set to satisfy the condition $f(\phi_2)=0$. There is no generic expression for it,\footnote{Alternatively, one can write \Eq{eq:f:sol} in the explicit form~\cite{Starobinsky:1986fx}
\begin{align*}
f\left(\varphi\right)=
\int_{\phi_1}^{\phi_2} \frac{\dd y}{\Mp} \int_y^{\phi_2} \frac{\dd x}{\Mp}\frac{1}{v(y)}\exp\left[\frac{1}{v(y)}- \frac{1}{v(x)}\right]\left[\theta(x-x_*)
- p_1\right]\, ,
\end{align*}
where $p_1$ is given by \Eq{stocha:hittproba:p1} and, in the configuration of \Fig{fig:sketch}, $\theta(x-x_*)=1$ when $x>x_*$ and $0$ otherwise.} but one can be more specific. First of all, as can be seen in \Fig{fig:NmeanIntegrationDomain}, $\bar{\phi}$ must be such that, when $f$ is evaluated at $\phi_2$, the integration domain of \Eq{eq:f:sol} possesses a positive part and a negative part, which are able to compensate for each other. This implies that $\bar{\phi}$ must lie between $\phi_1$ and $\phi_2$. A second generic condition comes from splitting the $x$-integral in \Eq{eq:f:sol} into $\int_{\phi_1}^{\varphi}\dd x = \int_{\phi_1}^{\phi_2} \dd x+ \int_{\phi_2}^{\varphi}\dd x$. The first integral vanishes because $f(\phi_2)=0$, which means that in order for $f$ to be symmetrical in $\phi_1\leftrightarrow \phi_2$, $\bar{\phi}(\phi_1,\phi_2)$ must satisfy this symmetry too, that is to say, $\bar{\phi}\left(\phi_1,\phi_2\right)=\bar{\phi}\left(\phi_2,\phi_1\right)$. Third, in the case where the potential is symmetric about a local maximum $\phi_\umax$ close to which inflation proceeds, the integrand in \Eq{eq:f:sol} is symmetric with respect to the first bisector in \Fig{fig:NmeanIntegrationDomain}. The two green triangles must therefore have the same surface, which readily leads to $\bar{\phi}=\phi_\umax$. Fourth, finally, in the case displayed in the right panel of \Fig{fig:sketch}, if $\phi_2$ is sufficiently large, we have established in section~\ref{sec:WallHittProba} that $p_2=0$ and the quantity we compute is the mean number of \efolds between $\phi_*$ and $\phi_1=\phi_\uend$. For explicitness, let us assume that $v^\prime>0$ (the same line of arguments applies in the case $v^\prime <0$). Inflation proceeds at $\phi<\phi_2$. In the domain of negative contribution in \Fig{fig:NmeanIntegrationDomain}, the argument of the exponential in \Eq{eq:f:sol} is positive. As a consequence, if $\bar{\phi}$ is finite and $\phi_2\rightarrow \infty$, the negative contribution to the integral is infinite while the positive one remains finite, which is impossible. In order to avoid this, one must then have $\bar{\phi} = \phi_2$. In practice, almost all cases boil down to one of the two previous ones and $\bar{\phi}$ is specified accordingly. Combining \Eqs{eq:meantime} and~(\ref{eq:f:sol}), one finally has\footnote{This is again in agreement with Eq. (35) of~\Ref{Starobinsky:1986fx} if $H$ is constant and $\phi_*=\bar{\phi}=0$, while $\phi_\uend =\infty$.}
\beq
\left\langle\mathcal{N}\right\rangle=\int^{\phi_*}_{\phi_\uend}\frac{\dd x}{\Mp}\int^{\bar{\phi}}_{x}\frac{\dd y}{\Mp}\frac{1}{v\left(y\right)}\exp \left[\frac{1}{v\left(y\right)}-\frac{1}{v\left(x\right)}\right]\, .
\label{eq:Nmean}
\eeq
This quantity is plotted for large and small field potentials in \Fig{fig:Nmean}, where it is compared with the results of a numerical integration of the Langevin equation~(\ref{eq:Langevin}) for a large number of realizations over which the mean value of $\mathcal{N}$ is computed. One can check that the agreement is excellent.
\paragraph{Classical Limit}
$ $\\
Let us now verify that the above formula boils down to the classical result~(\ref{eq:deltaNform:traj:class}) in some ``classical limit''. This can be done by performing a saddle-point approximation of the integrals appearing in \Eq{eq:Nmean}. Let us first work out the $y$-integral, that is to say, $\int^{\bar{\phi}}_x\dd y/v(y)\exp[1/v(y)]$. Since the integrand varies exponentially with the potential, the strategy is to evaluate it close to its maximum, \ie where the potential is minimum. The potential being maximal at $\bar{\phi}$ in most cases (see the discussion above), the integrand is clearly maximal\footnote{Strictly speaking, this is only true if the potential is a monotonous function of the field, but this is most often the case in the part of the potential that is relevant to the inflationary phase.} at $x$. Taylor expanding $1/v$ at first order around $x$, $1/v(y)\simeq 1/v(x)-v^\prime(x)/v^2(x)(y-x)$, one obtains, after integrating by parts,\footnote{Since $v(\bar{\phi})\gg v(x)$ and if $v$ is monotonous, one can also show that $\exp\left[-v^\prime(x)/v^2(x)(\bar{\phi}-x)\right]$ is exponentially vanishing and this term can be neglected.} $\int^{\bar{\phi}}_x\dd y/v(y)\exp\left[1/v(y)\right]\simeq v(x)/v^\prime(x)\exp\left[1/v(x)\right]$. Plugging back this expression into \Eq{eq:Nmean}, one finally obtains
\beq
\left.\left\langle\mathcal{N}\right\rangle\right\vert_\ucl= \int_{\phi_\uend}^{\phi_*}\frac{\dd x}{\Mp^2}\frac{v(x)}{v^\prime(x)}\, ,
\label{eq:stocha:meanN:classtraj}
\eeq
which exactly matches the classical result~(\ref{eq:deltaNform:traj:class}). The classical trajectory thus appears as a saddle-point limit of the mean stochastic trajectory, analogously to what happens \eg in the context of path integral calculations.

This calculation also allows us to identify under which conditions the classical limit is recovered. A priori, the Taylor expansion of $1/v$ can be trusted as long as the difference between $1/v(x)$ and $1/v(y)$ is not too large, say $\vert 1/v(y)-1/v(x)\vert < R$, where $R$ is some small number. If one uses the Taylor expansion of $1/v$ at first order, this means that $\vert y-x\vert<Rv^2/v'$. Requiring that the second order term of the Taylor expansion is small at the boundary of this domain yields the condition $\vert 2v-v^{\prime\prime}v^2/{v^\prime}^2\vert \ll 1$. For this reason, we define the classicality criterion
\beq
\eta_\ucl = \left\vert 2v-\frac{v^{\prime\prime}v^2}{{v^\prime}^2}\right\vert\, .
\label{eq:classicalcriterion:def}
\eeq
This quantity is displayed in the top axes in \Fig{fig:Nmean} and one can check that indeed, the classical trajectory is a good approximation to the mean stochastic one if and only if $\eta_\ucl\ll 1$. In the following, we will see that $\eta_\ucl$ is the relevant quantity to discuss the strength of the stochastic effects in general and in section~\ref{sec:discussion}, we will further discuss the physical implications of \Eq{eq:classicalcriterion:def}. 

For now, and for future use, let us give the first correction to the classical trajectory. This can be obtained going one order higher in the saddle-point approximation, that is to say, using a Taylor expansion of $1/v$ at second order. One obtains
\beq
\left.\left\langle \mathcal{N} \right\rangle\right\vert_{\eta_\ucl\ll 1}\simeq  \int_{\phi_\uend}^{\phi_*}\frac{\dd x}{\Mp^2}\frac{v(x)}{v^\prime(x)}\left[1+v\left(x\right)-\frac{v^{\prime\prime}\left(x\right) v^2\left(x\right)}{{v^{\prime}}^2\left(x\right)}+\cdots\right]\, ,
\label{eq:Nmean:vll1limit}
\eeq
where the dots stand for higher order terms. In the brackets of \Eq{eq:Nmean:vll1limit}, the two last terms stand for the first stochastic correction and one should not be surprised that, in general, when $\eta_\ucl$ is small, it is small. It is also interesting to notice that it is directly proportional to $\dd\epsilon_1/\dd N$. When $\epsilon_1$ increases as inflation proceeds, the stochastic leading correction is therefore positive and the stochastic effects tend to increase the realized number of $e$-folds, while when $\epsilon_1$ decreases as inflation proceeds, the correction is negative and the stochastic effects tend to decrease the number of $e$-folds, at least at linear order.
\subsubsection[Number of \efolds Variance]{Number of \texorpdfstring{$\bm{e}$}{$e$}-folds Variance}\label{sec:MeanNefDisp}
Let us now move on with the calculation of the dispersion in the number of $e$-folds, defined in \Eq{eq:deltacalN:def}. If one squares \Eq{eq:stild:stocha}, and takes the stochastic average of it, one obtains\footnote{This is again a non-trivial result since both the integrand and the upper bound of the integral appearing in \Eq{eq:stild:stocha} are stochastic quantities, but, as before, it can be shown in a rigorous way.}
\beq
\left\langle \mathcal{N}^2 \right\rangle = f^2\left(\phi_*\right)+2\Mp^2\left\langle \int_0^{\mathcal{N}} {f^{\prime}}^2\left[\phi\left(N\right)\right]v\left[\phi\left(N\right)\right]\dd N \right\rangle\, .
\label{eq:sto:nmeansquared:n2}
\eeq
In order to make use of the It\^o lemma, let then $g(\phi)$ be the function defined by
\beq
\label{eq:sto:n2:g:def}
g^{\prime\prime}\left(\phi\right)v\left(\phi\right)-g^\prime\left(\phi\right)\frac{v^\prime\left(\phi\right)}{v\left(\phi\right)} = -2{f^\prime}^2\left(\phi\right)v\left(\phi\right)\, ,
\eeq
where $f$ is the function defined in \Eq{eq:f:sol}. When  the It\^o lemma~(\ref{eq:ito}) is applied to $g\left(\phi\right)$, if one further sets $g\left(\phi_1\right)=g\left(\phi_2\right)=0$, one obtains
\bea
g\left(\phi_*\right)&=& 2\Mp^2 \left\langle \int_0^{\mathcal{N}} { f^{\prime}}^2\left[\phi\left(N\right)\right]v\left[\phi\left(N\right)\right]\dd N \right\rangle
\nonumber \\ &=&
\left\langle \mathcal{N}^2 \right\rangle -f^2\left(\phi_*\right)
=
\left\langle \mathcal{N}^2 \right\rangle -\left\langle \mathcal{N} \right\rangle^2\, ,
\eea
where the second equality is just a consequence of \Eq{eq:sto:nmeansquared:n2} and where the third equality is just a consequence of \Eq{eq:meantime}. Therefore, one just needs to solve \Eq{eq:sto:n2:g:def} with boundary conditions $g\left(\phi_1\right)=g\left(\phi_2\right)=0$ and to evaluate the resulting function at $\phi_*$ in order to obtain $\delta \mathcal{N}^2=g(\phi_*)$. The differential equation~(\ref{eq:sto:n2:g:def}) can formally be integrated, and one obtains
\beq
g\left(\phi_*\right)=2\int^{\phi_*}_{\phi_1}\dd x\int^{\bar{\phi}_2\left(\phi_1,\phi_2\right)}_x \dd y {f^\prime}^2\left(y\right) \exp\left[\frac{1}{v\left(y\right)}-\frac{1}{v\left(x\right)}\right]\, ,
\label{eq:g:sol}
\eeq
where $\bar{\phi}_2\left(\phi_1,\phi_2\right)$ is an integration constant that must be chosen in order to have $g\left(\phi_2\right)=0$. One can show that it satisfies the four properties listed in section~\ref{sec:meanN} for $\bar{\phi}$ and can therefore be specified in the same manner. With $\phi_1=\phi_\uend$, one then has
\beq
\delta \mathcal{N}^2=2\int_{\phi_*}^{\phi_\uend}\dd x\int_{\bar{\phi}_2}^x \dd y {f^\prime}^2\left(y\right) \exp\left[\frac{1}{v\left(y\right)}-\frac{1}{v\left(x\right)}\right]\, .
\label{eq:stocha:deltaN:sto:deltaN}
\eeq
\paragraph{Classical Limit}
$ $\\
As was done for the mean number of \efolds in section~\ref{sec:meanN}, let us derive the classical limit of \Eq{eq:stocha:deltaN:sto:deltaN}. Obviously, in the classical setup the trajectories are not stochastic and $\delta\mathcal{N}^2=0$, and what we are interested in here is the non-vanishing leading order contribution to $\delta\mathcal{N}^2$ in the limit $\eta_\ucl\ll 1$.  As before, the $y$-integral can be worked out with a saddle-point approximation, and one obtains\footnote{In the $\eta_\ucl\ll 1$ limit, $f$ is close to the classical trajectory~(\ref{eq:stocha:meanN:classtraj}) as shown in section~\ref{sec:meanN}, and one can take $f^\prime\left(y\right)\simeq v\left(y\right)/v^\prime\left(y\right)\Mp^{-2}$.} $\int^{\bar{\phi}_2}_x\dd y {f^\prime}^2\left(y\right)\exp\left[1/v(y)\right]\simeq v^4\left(x\right)/{v^\prime}^3\left(x\right)\exp\left[1/v\left(x\right)\right]/\Mp^4$. Plugging back this expression into \Eq{eq:stocha:deltaN:sto:deltaN}, one obtains
\beq
\left.\delta\mathcal{N}^2\right\vert_{\ucl}= \frac{2}{\Mp^4}\int_{\phi_\uend}^{\phi_*}\dd x\frac{v^4\left(x\right)}{{v^\prime}^3\left(x\right)}\, .
\label{eq:stocha:deltaN2:classapp}
\eeq
Finally, and for future use again, let us give the first correction to this classical limit. Going one order higher in the saddle-point approximation, one obtains
\beq
\left. \delta\mathcal{N}^2 \right\vert_{\eta_\ucl\ll 1}\simeq  \frac{2}{\Mp^4}\int_{\phi_\uend}^{\phi_*}\dd x\frac{v^4\left(x\right)}{{v^\prime}^3\left(x\right)}\left[1+6v\left(x\right)-5\frac{v^2\left(x\right)v^{\prime\prime}\left(x\right)}{{v^{\prime}}^2\left(x\right)}+\cdots\right]\, .
\label{eq:deltaN2:classlim}
\eeq
\subsubsection[Number of \efolds Skewness and Higher Moments]{Number of \texorpdfstring{$\bm{e}$}{$e$}-folds Skewness and Higher Moments}\label{sec:MeanNefskewness}
In the same manner, if one denotes the third moment of the distribution of number of \efolds by $m\left(\phi_*\right)=\delta\mathcal{N}^3$ defined in \Eq{eq:deltaN3:def}, one can show that $m(\phi)$ is the solution of the differential equation $m^{\prime\prime}-m^\prime v^\prime/v^2=-6f^\prime g^\prime$ that obeys $m(\phi_1)=m(\phi_2)=0$. As before, taking $\phi_1=\phi_\uend$ and $\phi_2=\phi_\infty$, one obtains
\beq
\delta\mathcal{N}^3=6\int_{\phi_*}^{\phi_\uend}\dd x\int_{\bar{\phi}_3}^x \dd y f^\prime(y) g^\prime(y)\exp\left[\frac{1}{v\left(y\right)}-\frac{1}{v\left(x\right)}
\label{eq:skewness:exact}
\right]
\eeq
where $\bar{\phi}_3$ can be set as $\bar{\phi}$. Similarly to above, making use of \Eqs{eq:Nmean:vll1limit} and~(\ref{eq:deltaN2:classlim}), a saddle-point approximation of this integral leads to the classical limit
\beq
\left.\delta\mathcal{N}^3\right\vert_{\eta_\ucl\ll 1}\simeq
\frac{12}{\Mp^6}\int_{\phi_\uend}^{\phi_*}\dd x\frac{v^7}{{v^\prime}^5}\left(1+14v -11\frac{v^2v^{\prime\prime}}{{v^\prime}^2}+\cdots\right)\, .
\label{eq:skewness:class}
\eeq

Let us finally explain how the same procedure can be iterated and higher order moments can be calculated. Let us denote the $p^\mathrm{th}$ momentum of the number of \efolds distribution by
\beq
\sigma_p\equiv\delta \mathcal{N}^p= \left\langle\left(\mathcal{N}-\left\langle\mathcal{N}\right\rangle\right)^p\right\rangle\, ,
\eeq
where, by convention, we set $\sigma_0=1$ and $\sigma_1=0$. As above, one can recursively show that $\sigma_p$ is the solution of the differential equation
\beq
\sigma_p^{\prime\prime}-\sigma_p^{\prime}\frac{v^\prime}{v^2}=-2pf^\prime\sigma_{p-1}^\prime-p\left(p-1\right){f^\prime}^2\sigma_{p-2}
\eeq
satisfying $\sigma_{p}\left(\phi_1\right)=\sigma_{p}\left(\phi_2\right)=0$. On then has
\beq
\sigma_p\left(\phi_*\right)=\int_{\phi_*}^{\phi_\uend}\dd x\int_{\bar{\phi}_{p}}^x\dd y\left[2pf^\prime\left(y\right)\sigma_{p-1}^\prime\left(y\right)+p\left(p-1\right){f^\prime}^2\left(y\right)\sigma_{p-2}\left(y\right)\right]\exp\left[\frac{1}{v\left(y\right)}-\frac{1}{v\left(x\right)}\right]\, .
\label{eq:sigmap}
\eeq
When $p=2$, this yields the variance~(\ref{eq:stocha:deltaN:sto:deltaN}); when $p=3$, the skewness~(\ref{eq:skewness:exact}) is obtained; when $p=4$, the kurtosis could be derived as well, and so on and so forth for any moment.
\section{Results}
\label{sec:Results}
We are now in a position where we can combine the intermediary results of the previous sections to give explicit, non-perturbative and fully generic expressions for the first correlation functions of curvature perturbations in stochastic inflation. We first derive the relevant formulas and their classical limits, before commenting on their physical implications in section~\ref{sec:discussion}.
\subsection{Power Spectrum}
\begin{figure}[t]
\begin{center}
\includegraphics[width=0.474\textwidth]{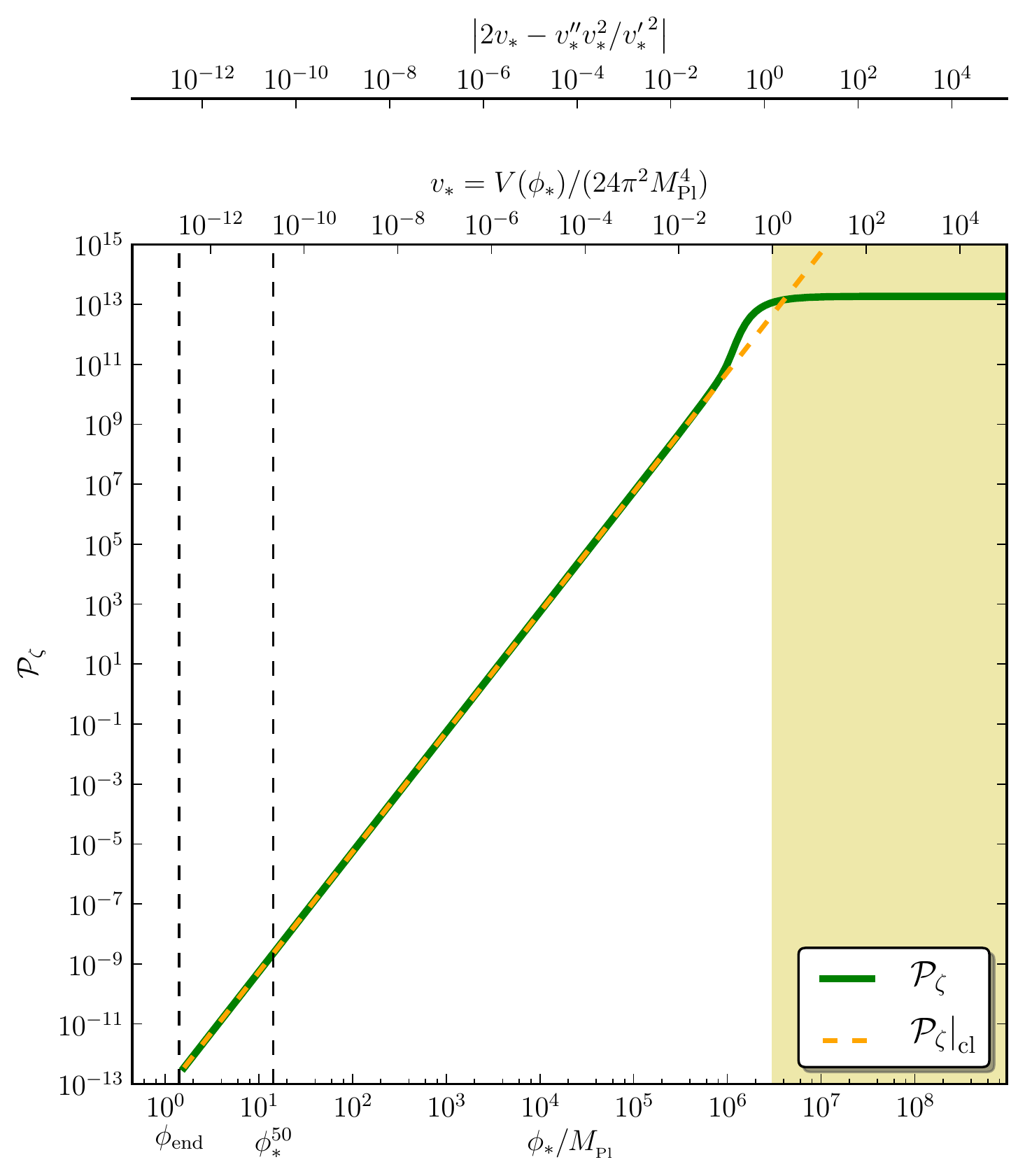}
\includegraphics[width=0.509\textwidth]{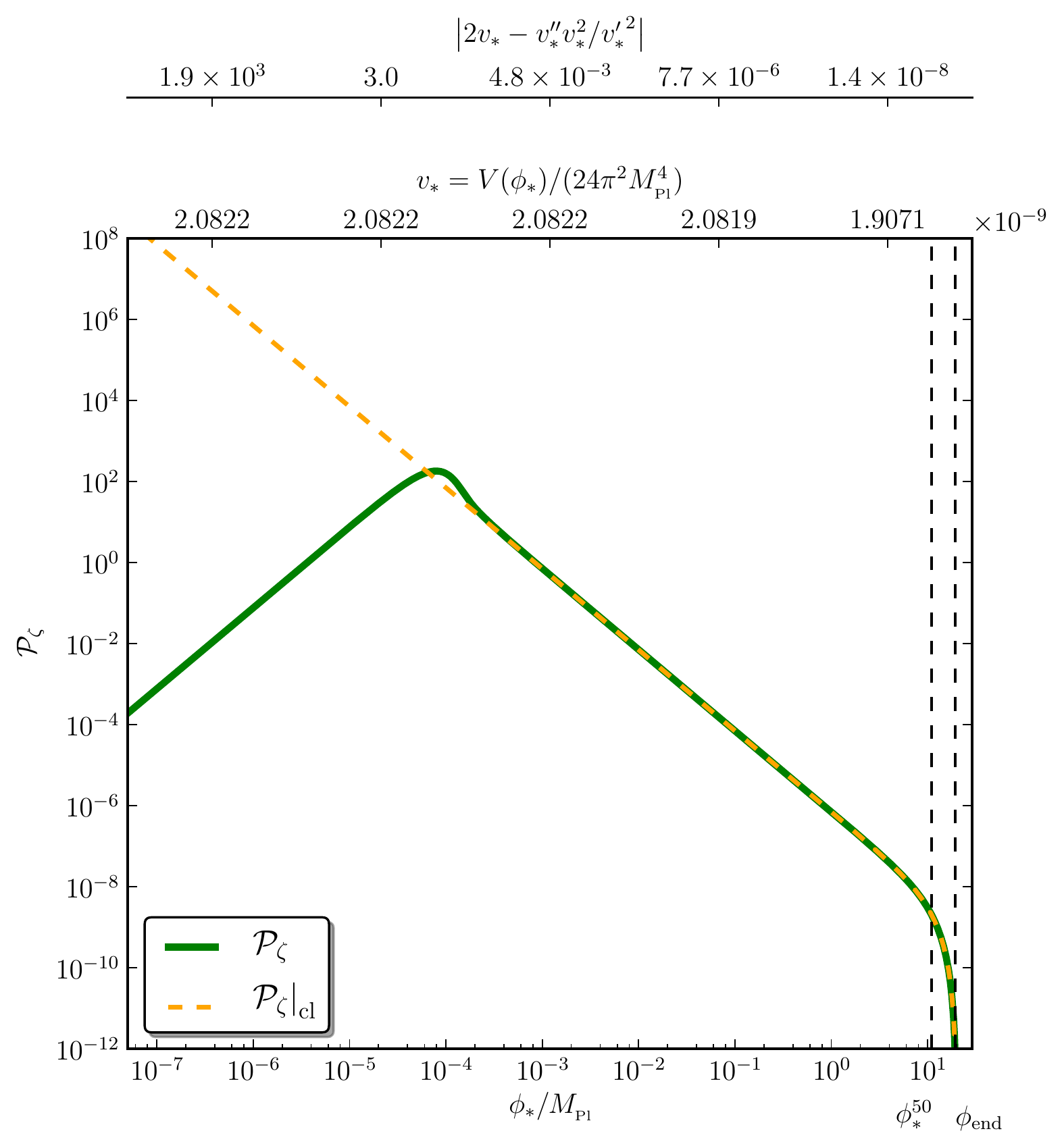}
\caption[Power spectrum]{Scalar power spectrum $\calP_\zeta $ for  the large field $V\propto\phi^2$ (left panel) and small field $V\propto 1-\phi^2/\mu^2$ (where $\mu=20\Mp$, right panel) potentials, as a function of $\phi_*$. The conventions are the same as in \Fig{fig:Nmean}. The green line corresponds to the analytical exact result~(\ref{eq:PS:fullstocha}), and the orange dashed line to the classical limit~(\ref{eq:stocha:PS:cl}).}
\label{fig:Pzeta}
\end{center}
\end{figure}
Following the programme we settled in section~\ref{sec:programme}, if one plugs \Eqs{eq:Nmean} and~(\ref{eq:stocha:deltaN:sto:deltaN}) into \Eq{eq:stocha:PS}, one obtains $\mathcal{P}_\zeta(\phi_*)=g^\prime(\phi_*)/f^\prime(\phi_*)$, that is,
\bea
\mathcal{P}_\zeta\left(\phi_*\right)&=&2
\left\lbrace \int_{\phi_*}^{\bar{\phi}}\frac{\dd x}{\Mp}\frac{1}{v\left(x\right)}\exp\left[\frac{1}{v\left(x\right)}-\frac{1}{v\left(\phi_*\right)}\right] \right\rbrace^{-1}
\times\nonumber\\ & &
\int_{\phi_*}^{\bar{\phi}_2}\frac{\dd x}{\Mp}\left\lbrace\int_{x}^{\phi_\infty} \frac{\dd y}{\Mp} \frac{1}{v\left(y\right)}\exp\left[\frac{1}{v\left(y\right)}-\frac{1}{v\left(x\right)}\right] \right\rbrace^2\exp\left[\frac{1}{v\left(x\right)}-\frac{1}{v\left(\phi_*\right)}\right]\, .
\label{eq:PS:fullstocha}
\eea
In this expression, $\mathcal{P}_\zeta(\phi_*)$ stands for the power spectrum calculated at a scale $k$ such that when it crosses the Hubble radius, the mean inflaton field value is $\phi_*$. This formula provides, for the first time, a complete expression of the curvature perturbations power spectrum calculated in stochastic inflation. It is plotted for large and small field potentials in \Fig{fig:Pzeta}.

From this, a generic expression for the spectral index can also be given. Since, at leading order in slow roll, $\partial/\partial\ln(k)\simeq-\partial\phi/\partial\langle\mathcal{N}\rangle\times\partial/\partial\phi$, one has
\beq
\nS=1-\frac{g^{\prime\prime}\left(\phi\right)}{f^\prime\left(\phi\right)g^\prime\left(\phi\right)}+\frac{f^{\prime\prime}\left(\phi\right)}{{f^\prime}^2\left(\phi\right)}\, .
\label{eq:stocha:ns:exact}
\eeq
Here, for conciseness, we do not expand this expression in terms of integrals of the potential, but it is straightforward to do so with \Eqs{eq:f:sol} and~(\ref{eq:g:sol}).
\paragraph{Classical Limit}
$ $\\
Before commenting further on the physical implications of the above result, let us make sure that in the classical limit, $\eta_\ucl\ll 1$, the standard formula is recovered. Combining \Eqs{eq:stocha:PS}, ~(\ref{eq:stocha:meanN:classtraj}) and~(\ref{eq:stocha:deltaN2:classapp}), one has
\beq
\left.\calP_\zeta\right\vert_{\ucl}\left(\phi_*\right)=\frac{2}{\Mp^2}\frac{v^3\left(\phi_*\right)}{{v^\prime}^2\left(\phi_*\right)}\, ,
\label{eq:stocha:PS:cl}
\eeq
which exactly matches the usual result~(\ref{eq:deltaNform:Pzeta}) at leading order in slow roll. This fully generalizes the work of \Ref{Fujita:2013cna}, where this agreement is shown but only in the case where the Hubble parameter varies linearly with $\phi$ and if the noise has constant amplitude. Here we have extended this result to any potential is single-field slow-roll inflation, and included dependence of the noise amplitude on the coarse grained field. In the same manner, making use of \Eqs{eq:stocha:meanN:classtraj}, (\ref{eq:stocha:deltaN2:classapp}) and~(\ref{eq:stocha:ns:exact}) together, one obtains $\left.\nS\right\vert_{\ucl}=1-\Mp^2\left[3\left(v^\prime/v\right)^2-2v^{\prime\prime}/v\right]$, which again matches the standard slow-roll result $\nS=1-2\epsilon_1-\epsilon_2$ where $\epsilon_2\equiv\dd\ln\epsilon_1/\dd N$ is the second slow-roll parameter, since at leading order in slow roll, one has $\epsilon_2=2\Mp^2({v^\prime}^2/v^2-v^{\prime\prime}/v)$, and $\epsilon_1$ has been given above.

Let us now derive the leading order corrections to these ``classical'' results. This can be done making use of \Eqs{eq:Nmean:vll1limit} and~(\ref{eq:deltaN2:classlim}) in \Eqs{eq:PS:fullstocha} and~(\ref{eq:stocha:ns:exact}). For the power spectrum, one obtains
\beq
\left.\mathcal{P}_\zeta\right\vert_{\eta_\ucl\ll 1}\left(\phi_*\right)\simeq\left.\calP_\zeta\right\vert_{\ucl}\left(\phi_*\right)\left[1+5v\left(\phi_*\right)-4\frac{v^2\left(\phi_*\right)v^{\prime\prime}\left(\phi_*\right)}{{v^\prime}^2\left(\phi_*\right)}\right]\, ,
\label{eq:PS:vll1}
\eeq
while for the spectral index one gets
\beq
\left.\nS\right\vert_{\eta_\ucl\ll 1}\left(\phi_*\right)\simeq\left.\nS\right\vert_{\ucl}\left(\phi_*\right)+\Mp^2\left[3v^{\prime\prime}\left(\phi_*\right)-2\frac{{v^\prime}^2\left(\phi_*\right)}{v\left(\phi_*\right)}-6\frac{{v^{\prime\prime}}^2\left(\phi_*\right)v\left(\phi_*\right)}{{v^\prime}^2\left(\phi_*\right)}+4\frac{v\left(\phi_*\right)v^{\prime\prime\prime}\left(\phi_*\right)}{v^\prime\left(\phi_*\right)}\right]\, .
\label{eq:ns:vll1}
\eeq
\subsection{Non-Gaussianity and Higher Moments}
The local non-Gaussianity parameter can be calculated in the same manner, and \Eq{eq:stocha:fnl} gives rise to
\beq
\fnl=\frac{5}{72}\left(\frac{m^{\prime\prime}}{{g^\prime}^2}-\frac{f^{\prime\prime}m^\prime}{f^\prime{g^\prime}^2}\right)\, .
\label{eq:fnl:exact}
\eeq
For conciseness, this expression is not expanded in terms of integrals of the potential, but it is straightforward to do so with \Eqs{eq:f:sol}, (\ref{eq:g:sol}) and~(\ref{eq:skewness:exact}). 

Here also, we need to make sure that in the classical limit, the standard result is recovered. Combining \Eqs{eq:fnl:exact}, ~(\ref{eq:Nmean:vll1limit}), (\ref{eq:deltaN2:classlim}) and~(\ref{eq:skewness:class}), one obtains
\beq
\left.\fnl\right\vert_{\eta_\ucl\ll 1}=\frac{5}{24}\Mp^2\left[6\frac{{v^\prime}^2}{v^2}-4\frac{v^{\prime\prime}}{v}+v\left(25\frac{{v^{\prime}}^2}{v^2}-34\frac{v^{\prime\prime}}{v}-10\frac{v^{\prime\prime\prime}}{v^\prime}+24\frac{{v^{\prime\prime}}^2}{{v^\prime}^2}\right)+\mathcal{O}\left(v^2\right)\right]\ .
\label{eq:fnl:classappr}
\eeq
The first two terms in the brackets match the usual result~\cite{Maldacena:2002vr}. In contrast, it is important to stress that within the usual $\delta N$ formalism, the standard result cannot be obtained because of the intrinsic non-Gaussianity of the fields at Hubble exit~\cite{Maldacena:2002vr, Allen:2005ye}. Such effects are automatically taken into account in our formalism, which readily gives rise to the correct formula. 

Obviously, one can go on and calculate any higher order correlation function with \Eq{eq:sigmap}. However, with the power spectrum and non-Gaussianity local parameter at hand, we already are in a position where we can draw important physical conclusions.
\subsection{Discussion}
\label{sec:discussion}
A first important consequence of \Eqs{eq:PS:fullstocha} and~(\ref{eq:fnl:exact}) is the correctness of their classical limits. They show the validity of our computational programme for calculating correlation functions in general. This may be particularly useful for investigating other cases than single-field slow-roll inflation, especially when the standard procedure is difficult to follow. Indeed, our method can easily be numerically implemented, and it could then be applied to more complicated scenarios such as multi-field inflation where it has been shown~\cite{Watanabe:2011sm} that the $\delta N$ formalism retains reliable, modified kinetic terms where the stochastic inflation formalism has been generalized~\cite{Chen:2006hs, Helmer:2006tz, Tolley:2008na, Lorenz:2010vf}, \etc{} In particular, it is well suited to situations where stochastic effects dominate the inflationary dynamics in some parts of the potential~\cite{Martin:2011ib, Levasseur:2013tja,Fujita:2014tja} and where one must take the stochastic effects into account.

Let us mention that within the CMB observable window, corrections to the classical results are always small, since one has 
\beq
\eta_\ucl\simeq \calP_\zeta\left(\epsilon_1+\frac{\epsilon_2}{4}\right)\, .
\eeq
More precisely, \Eqs{eq:PS:vll1} and~(\ref{eq:fnl:classappr}) can be recast as $\left.\mathcal{P}_\zeta\right\vert_{\eta_\ucl\ll 1}\simeq  \left.\calP_\zeta\right\vert_{\ucl}\left[1+\left.\calP_\zeta\right\vert_{\ucl}(\epsilon_1+\epsilon_2)\right]$ and $\left.\fnl\right\vert_{\eta_\ucl\ll 1}\simeq  \left.\fnl\right\vert_{\ucl}- \frac{5}{12}\left.\mathcal{P}_\zeta\right\vert_{\ucl}\left(38\epsilon_1^2+\frac{51}{4}\epsilon_2\epsilon_1+\frac{9}{8}\epsilon_2\epsilon_3-\frac{59}{8}\epsilon_2^2\right)$, where $\epsilon_3\equiv\dd\ln\epsilon_2/\dd N$ is the third slow-roll parameter. For the scales of astrophysical interest today, in standard single-field slow-roll inflation, these corrections are therefore tiny.

However, even if the stochastic effects within the CMB observable window need to be small, let us stress that the location of the observable window along the inflationary potential can be largely affected. This notably happens when the potential has a flat region between the location where the observed modes exit the Hubble radius and the end of inflation, as is the case \eg in hybrid inflation or in potentials with flat inflection points.

Another point to note is that, contrary to what one may have expected, the corrections we obtained are not controlled by the ratio $\Delta\phi_{\mathrm{qu}}/\Delta\phi_\ucl$ extensively used in the literature, but by the classicality criterion $\eta_\ucl$ derived in \Eq{eq:classicalcriterion:def}. This has two main consequences. 

First, $\eta_\ucl$ has dimension $v$, which means that it is Planck suppressed.\footnote{This remark also sheds some new light on the old debate~\cite{Linde:2005ht, Smolin:1979ca, Bardeen:1983st} whether quantum gravitational corrections should affect inflationary predictions through powers of $\phi/\Mp$ or $V/\Mp^4$. This analysis reveals $V/\Mp^4$ corrections only, regardless of the value of $\phi/\Mp$.} This makes sense, since, as noted in section~\ref{sec:statDistrib}, some of the corrections we obtained physically correspond to the self- and gravitational interactions of the inflaton field.\footnote{For this reason, one may think that performing the calculation in Fourier space as we did does not allow us to properly account for self-interaction effects and that a real space calculation should be carried out instead. However, since the separate Universe approximation is exponentially well verified on large scales, this is not the case. Making use of the same formalism as in \Ref{Starobinsky:1994bd}, we have indeed explicitly checked that performing the calculation in real space leads to  the same results as the ones presented here.} This is why it can be useful to compare our results with loop calculations performed in the literature by means of other techniques. In particular, the self-loop correction to the power spectrum is derived in \Ref{Seery:2007we}, and graviton loop corrections are obtained in \Ref{Dimastrogiovanni:2008af} (for a nice review, see also \Ref{Seery:2010kh}). A diagrammatic approach based on the $\delta N$ formalism is also presented in \Ref{Byrnes:2007tm} where the power spectrum and the bispectrum are calculated up to two loops. In all these cases, the obtained corrections are of the form $\calP_\zeta^{1\mathrm{-loop}}=\calP_\zeta^\mathrm{tree}(1+\alpha\calP_\zeta^\mathrm{tree}\epsilon^2 N)$. Here, $\alpha$ is a numerical factor of order one that depends on the kind of loops one considers, and $\epsilon^2$ stands for second order combinations of slow-roll parameters. When the number of \efolds $N$ is of the order $1/\epsilon$, this is exactly the kind of leading corrections we obtained. This feature is therefore somewhat generic. Obviously, it remains to understand which loops exactly our approach allows one to calculate, and how our results relate to the above mentioned ones. We leave it for future work.
\begin{table}[t]
\begin{center}
\begin{tabular}{{|l||c|c|}}
\hline
\textbf{Potential type} & $\boldsymbol{v(\phi)}$ & $\boldsymbol{\eta_\ucl}$\\
\hline
Large field & $\propto\phi^p$ & $\left(1+\frac{1}{p}\right)v$ \\
\hline
Hilltop & $ v_0\left[1-\left(\frac{\phi}{\mu}\right)^p\right]$ & $\frac{v_0}{p}\left(\frac{\mu}{\phi}\right)^p$ \\
\hline
Polynomial plateau & $ v_\infty\left[1-\left(\frac{\phi}{\mu}\right)^{-p}\right]$ & $\frac{v_\infty}{p}\left(\frac{\phi}{\mu}\right)^p$\\
\hline
Exponential plateau & $ v_\infty\left[1-\alpha\exp\left(-\frac{\phi}{\mu}\right)\right]$ & $\frac{v_\infty}{\alpha}\exp\left(\frac{\phi}{\mu}\right)$ \\
\hline
Inflection point & $v_0\frac{n\left(n-1\right)}{\left(n-1\right)^2}\left[\left(\frac{\phi}{\phi_0}\right)^2-\frac{4}{n}\left(\frac{\phi}{\phi_0}\right)^n+\frac{1}{n-1}\left(\frac{\phi}{\phi_0}\right)^{2n-2}\right]$ & $\frac{v_0}{n\left(n-1\right)}\left\vert\frac{\phi}{\phi_0}-1\right\vert^{-3}$\\
\hline
\end{tabular}
\end{center}
\caption[Classicality Criterion for a few Potentials] {Classicality criterion $\eta_\ucl$ defined in \Eq{eq:classicalcriterion:def} for a few types of inflationary potentials. Except for ``large field'', the expression given for $\eta_\ucl$ is valid close to the flat point of the potential.} \label{tab:etaclass} 
\end{table}

Second, $\eta_\ucl$ contains $1/v'^2$ terms. This means that, even if $v$ needs to be very small,\footnote{Since $v$ can only decrease, $v<10^{-10}$ for all observable modes.} if the potential is sufficiently flat, $\eta_\ucl$ may be large. In table~\ref{tab:etaclass}, we have summarized the shape of $\eta_\ucl$ for different prototypical inflationary potentials. For large field potentials, $\eta_\ucl$ is directly proportional to $v$. This is why, in the left panels of \Figs{fig:Nmean} and~\ref{fig:Pzeta}, departure of the stochastic results from the standard formulas occur only when $v\gg 1$, in a regime where our calculation cannot be trusted anyway. However, for potentials with flat points, different results are obtained. If the flat point is of the hilltop type, $\eta_\ucl$ diverges at the maximum of the potential. This is why, in the right panels of \Figs{fig:Nmean} and~\ref{fig:Pzeta}, even if $v$ saturates to a small maximal value, the stochastic result differs from the classical one close to the maximum of the potential. However, this happens many \efolds before the scales probed in the CMB cross the Hubble radius, that is to say, at extremely large, non-observable scales. The same conclusion holds for plateau potentials (either of the polynomial or exponential type) where stochastic effects lead to non-trivial modifications in far, non observable regions of the plateau. On the other hand, if the potential has a flat inflection point, $\eta_\ucl$ can be large at intermediate wavelengths, too small to lie in the CMB observable window but still of astrophysical interest. This could have important consequences in possible non-linear effects at those small scales, such as the formation of primordial black holes (PBHs)~\cite{Green:2004wb}. In such models, the production of PBHs is calculated making use of the standard classical formulas for the amount of scalar perturbations. However, we have shown that in such regimes, stochastic effects largely modify its value. An important question is therefore how this changes the production of PBHs in these models. In particular, it is interesting to notice that if the potential is concave ($v^{\prime\prime} < 0$), which is the case favored by observations~\cite{Martin:2013tda, Martin:2013nzq, Martin:2014lra}, the leading correction in \Eq{eq:PS:vll1} is an enhancement of the power spectrum amplitude. However, as can be seen in the right panel of \Fig{fig:Pzeta}, as soon as one leaves the perturbative regime, this is replaced by the opposite trend: at the flat point, the classical result accounts for a diverging power spectrum while the stochastic effects make it finite. If generic, this effect may be important for the calculation of PBHs formation, and we plan to address this issue in a future publication.
\section{Conclusion}\label{sec:stocha:powerspectrum:discussion}
\label{sec:Conclusion}
Let us now summarize our main results. Making use of the $\delta N$ formalism, we have shown how curvature perturbations can be related to fluctuations in the realized amount of inflationary \efolds in stochastic inflation trajectories. We have then applied ``first passage time analysis'' techniques to derive all the statistical moments of the number of $e$-folds, hence all scalar correlation functions in stochastic inflation.

We have shown that the standard results can be recovered as saddle-point limits of the full expressions. The situation is therefore analogous to, \eg, path integral calculations. A new simple classicality criterion has been derived, which should replace the common estimate based on the ratio between the mean quantum kick and the classical drift during one $e$-fold. It shows that quantum corrections to inflationary observables are Planck suppressed in general (that is to say, they are proportional to $V/\Mp^4$), but can be large if the potential is flat enough, even at sub-Planckian scales. For simple inflationary models where $\vert\dd V/\dd \phi\vert/V$ increases monotonously as inflation proceeds, the corresponding effects play a non-trivial role only at extremely large, non-observable scales. However, models containing a flat point in the potential between the Hubble exit location of the modes currently observed in the CMB and the end of inflation behave differently. First, the stochastic effects change the mean total number of inflationary \efolds and can therefore largely modify the location of the observational window along the inflationary potential. Second, the amount of scalar perturbations produced around the flat point is strongly modified by stochastic effects. This may be crucially important for a number of non-linear effects computed at these small scales, such as the formation of PBHs, or non-Gaussianity. 

Together with the case of tensor perturbations, which we have not addressed in this paper, we plan to study these issues in future publications.
\section*{Acknowledgements}
It is a pleasure to thank Eiichiro Komatsu, Giovanni Marozzi, J\'er\^ome Martin, Gianmassimo Tasinato, David Wands and Jun'ichi Yokoyama for very useful comments and enjoyable discussions. We also thank Oliver Jannsen for pointing out some small typos in the first archived version of our paper. VV's work is supported by STFC grant ST/L005573/1. A.S. was partially supported by the RFBR Grant No. 14-02-00894. His visit to the Utrecht University was supported by the Delta ITP Grant BN.000396.1. He thanks Profs. S. Vandoren and T. Prokopec and Dr. J. van Zee for hospitality during the visit.
\appendix
\section[Why must we use the number of \texorpdfstring{$\bm{e}$}{$e$}-folds in the Langevin equation?]{Why must we use the number of \texorpdfstring{$\bm{e}$}{$e$}-folds in the Langevin equation?}
\label{sec:app:whyN}
In section~\ref{sec:statDistrib}, we have made explicit that different choices of time variable in the Langevin equation account for different stochastic processes. In this appendix, we explain why the correct time variable to work with is the number of \efolds $N$, elaborating on already existing results. We first present a generic argument, based on a perturbation of the background equations, before turning to an explicit comparison of stochastic inflation and QFT predictions, in order to identify the correct time variable.
\subsection{Perturbations Equation derived from the Background Equation}\label{sec:stocha:whyN:VaryBackgroundEq}
A heuristic derivation~\cite{Starobinsky:1986fx} of the Langevin equation relies on splitting the full quantum inflaton field into a coarse grained, large scales part $\varphi$, and a short-wavelength component $\phi_>$, and on performing an expansion of the equation of motion in $\phi_>$. As suggested in \Ref{Finelli:2011gd}, the correct time variable should therefore be the one such that the equations for the perturbations, which must be established at the action level, can correctly be obtained from varying the equation of motion for the background itself, when written in terms of this time variable. In this section, we establish that this condition selects out $N$ as the time variable.

In the case where inflation is driven by a single scalar field $\phi$, the action we start from is given by
\beq
\mathcal{S}=\int\dd^4x\sqrt{-g}\left[\frac{\Mp^2}{2}R-\frac{1}{2}g^{\mu\nu}\partial_\mu\phi\partial_\nu\phi-V\left(\phi\right)\right]\, .
\label{eq:stocha:whyN:action}
\eeq
From this action (and this action only), we first want to derive equations of motion for the scalar perturbations that can be compared with what will be obtained below from varying the background equation of motion itself. To make our point even more convincing, we go up to second order in the perturbations. This is why we expand the background fields $\lbrace \phi,g_{\mu\nu} \rbrace$ at second order in the scalar perturbations.\footnote{In this discussion, vector and tensor perturbations are irrelevant, which is why they are not taken into account.} When the time variable in the metric is the conformal time $\eta$, one has
\bea
\phi\left(\eta,\vec{x}\right)&=&\phi^{(0)}\left(\eta\right)+\phi^{(1)}\left(\eta,\vec{x}\right)+\frac12\phi^{(2)}\left(\eta,\vec{x}\right)\, ,\nonumber\\
g_{00}&=&a^2\left[-1-2\alpha^{(1)}-\alpha^{(2)}\right]\, ,
\quad\quad\quad\quad
g_{i0}=-a^2\left[\partial_i B^{(1)}+\frac12\partial_i B^{(2)}\right]\, ,\\
g_{ij}&=&a^2\left\lbrace \delta_{ij}\left[1-2\psi^{(1)}-\psi^{(2)}\right]
+2\partial_i\partial_j\left[E^{(1)}+\frac12 E^{(2)}\right]\right\rbrace\, .\nonumber
\eea
The degrees of freedom introduced above are partially redundant and in absence of anisotropic stress, the scalar sector can be described in terms of a single gauge invariant variable. One possible choice is the Mukhanov-Sasaki variable~\cite{Mukhanov:1981xt, Kodama:1985bj, Mukhanov:1988jd} $v$, which can be defined, order by order, as the scalar field fluctuation $\phi^{(n)}$ on uniform curvature hypersurfaces~\cite{Malik:2005cy}. To first and second orders, after a lengthy but straightforward calculation, one obtains~\cite{Malik:2003mv}
\bea
\label{eq:stoch:MukhanovSasaki:def:1}
v^{(1)}&=&\phi^{(1)}+\frac{\left(\phi^{(0)}\right)^\prime}{\mathcal{H}}\psi^{(1)}\, ,\\
v^{(2)}&=&\phi^{(2)}+\frac{\left(\phi^{(0)}\right)^\prime}{\mathcal{H}}\psi^{(2)}
+\left(\frac{\psi_1}{\mathcal{H}}\right)^2\left[\left(\phi^{(0)}\right)^{\prime\prime}+2\mathcal{H}\left(\phi^{(0)}\right)^\prime-\frac{\mathcal{H}^\prime}{\mathcal{H}}\left(\phi^{(0)}\right)^\prime\right]
\nonumber\\ & &
+2\frac{\left(\phi^{(0)}\right)^\prime}{\mathcal{H}^2}\psi_1^\prime\psi_1+2\frac{\psi_1}{\mathcal{H}}\left(\phi^{(1)}\right)^\prime\, .
\label{eq:stoch:MukhanovSasaki:def:2}
\eea

From varying the expanded action, one can derive an equation of motion for the scalar perturbations, and in particular for a gauge invariant combination of them, say the Mukhanov-Sasaki variable. In this section, we want to compare this action based equation of motion for $v$ with an equation of motion for the perturbation in $\phi$ coming from varying the background Klein-Gordon equation. It is therefore important to work in a gauge where these two quantities, $v$ and the perturbation in $\phi$, are identical. By definition of the Mukhanov-Sasaki variable, this is the case in the uniform curvature gauge, where
\beq
\label{eq:MukhanovVar:UCG}
v^{(n)}=\phi^{(n)}
\eeq
to all orders. In this gauge, one notably has $\psi=0$ to all orders [from \Eqs{eq:stoch:MukhanovSasaki:def:1}, (\ref{eq:stoch:MukhanovSasaki:def:2}) and~(\ref{eq:MukhanovVar:UCG}), it is clear that this is at least the case up to second order].

The equation of motion for the scalar perturbations $\phi^{(1)}$ and $\phi^{(2)}$ is therefore given by the one for $v^{(1)}$ and $v^{(2)}$ in this gauge. At leading order in the slow-roll approximation, and in the long-wavelength limit, they read\footnote{In spite of the complexity of the field equations at second order, see \eg \Ref{Noh:2003yg}, in the long-wavelength limit, it is sufficient~\cite{Wands:2000dp} to use the local conservation of energy-momentum to establish \Eqs{eq:stoch:pert1:eom} and~(\ref{eq:stoch:pert2:eom}). Because this is not the main subject of this discussion, the corresponding calculations are not reproduced here but they can be found in \Refs{Wands:2000dp, Malik:2005cy}.}
\bea
\label{eq:stoch:pert1:eom}
3H\dot{\phi}^{(1)}+\left(V^{\prime\prime}-\frac{{V^\prime}^2}{3H^2\Mp^2}\right)\phi^{(1)}&=&0\, ,\\
3H\dot{\phi}^{(2)}+\left(V^{\prime\prime}-\frac{{V^\prime}^2}{3H^2\Mp^2}\right)\phi^{(2)}&=&-\frac{1}{2}\left(V^{\prime\prime\prime}-\frac{V^\prime V^{\prime\prime}}{H^2\Mp^2}+\frac{2{V^\prime}^3}{9H^4\Mp^4}\right){\phi^{(1)}}^2\, .
\label{eq:stoch:pert2:eom}
\eea

We now need to compare these equations with the ones that arise when varying the equation of motion for the background, and to find for which time variable they match.
\paragraph{If \texorpdfstring{$\bm{t}$}{$t$} is used}
$ $\\
When cosmic time $t$ is used, the leading order of the slow-roll approximation for the Klein-Gordon equation for the background is given by
\beq
\frac{\dd \phi}{\dd t}=-\frac{V^\prime}{3H\left(\phi\right)}\, ,
\eeq
where we take $H^2\simeq V/(3\Mp^2)$ at leading order in slow roll. When plugging $\phi=\phi^{(0)}+\phi^{(1)}+\phi^{(2)}$ in this equation, one obtains at first and second order in the perturbations
\bea
3H\tilde{\dot{\phi}}^{(1)}+\left(V^{\prime\prime}-\frac{{V^\prime}^2}{{\color{red}{6}} H^2\Mp^2}\right)\tilde{\phi}^{(1)}&=&0\, ,\\
3H\tilde{\dot{\phi}}^{(2)}+\left(V^{\prime\prime}-\frac{{V^\prime}^2}{{\color{red}{6}} H^2\Mp^2}\right)\tilde{\phi}^{(2)}&=&-\frac{1}{2}\left(V^{\prime\prime\prime}-\frac{V^\prime V^{\prime\prime}}{{\color{red}{2}}H^2\Mp^2}+\frac{{V^\prime}^3}{{\color{red}{12}}H^4\Mp^4}\right){{}{\tilde{\phi}}^{(1)}}^{2}\, .
\eea
One should stress that these equations do not apply to $\phi^{(1)}$ and $\phi^{(2)}$ since they are different from \Eqs{eq:stoch:pert1:eom} and~(\ref{eq:stoch:pert2:eom}), which is why we use the notation $\tilde{\phi}^{(1,2)}$ instead of $\phi^{(1,2)}$. The differences with \Eqs{eq:stoch:pert1:eom} and~(\ref{eq:stoch:pert2:eom}) are displayed in red. One can see that several factors do not match. This is because in general, the equations for the perturbations must be derived from the action itself and cannot be obtained by simply varying the equation of motion for the background.
\paragraph{If \texorpdfstring{$\bm{\dd s=H^pa^q \dd t}$}{$\dd s=H^p\left(\phi\right) \dd t$} is used}
$ $\\
For this reason, let us look for a time variable $s$ which is such that the equations for the perturbations arise from varying the equation of motion of the background when written in terms of $s$. Let us assume that $s$ is related to $t$ thanks to a relation of the form
\beq
\dd s=H^p\left(\phi\right)a^q\left(\phi\right) \dd t\, ,
\eeq
where $p$ and $q$ are power indices that we try to determine. For example, when $p=0$ and $q=0$, $s$ is the cosmic time $t$, when $p=1$ and $q=0$, $s$ is the number of \efolds $N$, while when $p=0$ and $q=-1$, $s$ is the conformal time $\eta$. In terms of $s$, the equation of motion for the background is given by
\beq
\frac{\dd\phi}{\dd s}=-\frac{V^\prime}{3H^{p+1}\left(\phi\right)}\, ,
\eeq
where again we take $H^2\simeq V/(3\Mp^2)$ at leading order in slow roll. When plugging in $\phi=\phi^{(0)}+\phi^{(1)}+\phi^{(2)}$, one obtains at first and second order in the perturbations
\bea
& &\kern-1.5em 3H\dot{\tilde{\phi}}^{(1)}+\left(V^{\prime\prime}-{\color{red}{\frac{p+1}{6}}}\frac{{V^\prime}^2}{H^2\Mp^2} +3{\color{red}{q}}H^2\right)\tilde{\phi}^{(1)}= 0\, ,\\ & &\kern-1.5em
3H\dot{\tilde{\phi}}^{(2)}+\left(V^{\prime\prime}-{\color{red}{\frac{p+1}{6}}} \frac{{V^\prime}^2}{ H^2\Mp^2}+3{\color{red}{q}}H^2\right)\tilde{\phi}^{(2)}=
\nonumber\\ & &
-\frac{1}{2}\left[V^{\prime\prime\prime}-{\color{red}{\frac{p+1}{2}}}\frac{V^\prime V^{\prime\prime}}{H^2\Mp^2}+{\color{red}{\frac{\left(p+1\right)\left(p+3\right)}{36}}}\frac{{V^\prime}^3}{H^4\Mp^4}+3{\color{red}{q}}\frac{H^2V^{\prime\prime}}{V^\prime} -{\color{red}{pq}}\frac{V^\prime}{\Mp^2}-18{\color{red}{q}}\frac{H^4}{V^\prime}\right]{{}{\tilde{\phi}}^{(1)}}^{2}\, .\nonumber\\& &
\eea
Again, these equations do not apply to $\phi^{(1)}$ and $\phi^{(2)}$ in general, since the correct ones are given by \Eqs{eq:stoch:pert1:eom} and~(\ref{eq:stoch:pert2:eom}) which is why we use the notation $\tilde{\phi}^{(1,2)}$. The differences between these two sets of equations are displayed in red. In order for the above to match \Eqs{eq:stoch:pert1:eom} and~(\ref{eq:stoch:pert2:eom}), one must have $q=0$ and $(p+1)/6=1/3$, which gives $p=1$, $(p+1)/2=1$, which also gives $p=1$, and $(p+1)(p+3)/36=2/9$, which gives $p=1$ or $p=-5$. As a conclusion, with $p=1$ and $q=0$ only, the equations for the perturbations (from what is shown here, up to second order in perturbation theory) can be seen as if they were derived from varying the equation of motion for the background. This choice corresponds to the number of \efolds $N$. 
\subsection{Stochastic Inflation and Quantum Field Theory on Curved Space-Times}
\label{sec:sto:StoAndQFT}
To go beyond this generic argument, one can explicitly show~\cite{Finelli:2008zg, Finelli:2010sh} that $N$ is the time variable which allows one to consistently connect stochastic inflation with results from QFT on curved space-times. For example, let us consider the leading order of the fluctuations $\delta\phi=\varphi-\phi_\ucl$ in the coarse grained inflaton field about its classical background value $\phi_\ucl$. By ``classical'', recall that we mean that $\phi_\ucl$ is the solution of the equation of motion without the noise term. We want to compute the mean square value of $\delta\phi$ and compare what we obtain with results coming from QFT calculations. For example, in \Ref{Finelli:2003bp}, with a renormalization obtained by employing the adiabatic subtraction prescription on inflationary backgrounds, it is shown that in quadratic inflation where $V=m^2\phi^2/2$, if $\delta\phi=0$ at time $t_0$, one has at leading order $\lbrace$see Eq.~(48) of \Ref{Finelli:2003bp}$\rbrace$
\beq
\label{eq:stoch:deltaphi1Squared:lfi}
\left\langle \left(\phi-\phi_\ucl\right)^2 \right\rangle = \frac{H_0^6-H^6}{8\pi^2m^2H^2}\, ,
\eeq
where $H$ means $H\left(\phi_\ucl\right)$ and $H_0$ means $H$ evaluated at time $t_0$. In the same manner, in \Ref{Marozzi:2006ky}, it was shown that in power-law inflation where $a(t)\propto t^{p}$ with $p\gg 1$, the same quantity is given by $\lbrace$see Eq.~(29) of \Ref{Marozzi:2006ky}$\rbrace$
\beq\label{eq:stoch:deltaphi1Squared:pli}
\left\langle \left(\phi-\phi_\ucl\right)^2 \right\rangle = \frac{p}{8\pi^2}\left(H_0^2-H^2\right)\, .
\eeq
Let us see how these results can be derived in the stochastic inflationary framework. We start from the Langevin equation~(\ref{eq:Langevin}) that we write
\beq
\frac{\dd\varphi}{\dd N}=-2\Mp^2 \frac{H^\prime}{H}+\frac{H}{2\pi}\xi\left(N\right)\, ,
\label{eq:Langevin:N:H}
\eeq
where we have used $H^2\simeq V/(3\Mp^2)$ and where we recall that a prime denotes a derivative with respect to the inflaton field. Since $\phi_\ucl$ is the solution of the above equation without the noise term, the noise term can be considered as a perturbation captured in $\delta\phi$. After expanding \Eq{eq:Langevin:N:H} in powers of $\delta\phi$, one gets for the leading order $\delta\phi^{(1)}$
\beq
\frac{\dd\delta\phi^{(1)}}{\dd N}+2\Mp^2\left(\frac{H^\prime}{H}\right)^\prime\delta\phi^{(1)}=\frac{H}{2\pi}\xi\, .
\label{eq:stocha:equadiff:deltaphi1}
\eeq
Multiplying this equation by $\delta\phi^{(1)}$ and taking the stochastic average $\langle \cdot\rangle$ leads to
\beq
\frac{\dd \left\langle \delta{\phi^{(1)}}^2 \right\rangle }{\dd N}+4\Mp^2\left(\frac{H^\prime}{H}\right)^\prime\left\langle \delta{\phi^{(1)}}^2 \right\rangle=\frac{H}{\pi}\left\langle \xi\delta{\phi^{(1)}} \right\rangle\, .
\eeq
In order to obtain a differential equation for $\langle\delta{\phi^{(1)}}^2 \rangle $ only, one needs to evaluate the right hand side of the previous equation. This can be done as follows. Letting $\delta{\phi^{(1)}}=0$ at time $N_0$, a formal solution of \Eq{eq:stocha:equadiff:deltaphi1} is given by
\beq
\delta{\phi^{(1)}}=\exp\left[-2\Mp^2\int_{N_0}^N\left(\frac{H^\prime}{H}\right)^\prime\dd n \right]\int_{N_0}^N\left\lbrace \frac{H}{2\pi}\xi\left(n\right)\exp\left[2\Mp^2\int_{N_0}^n\left(\frac{H^\prime}{H}\right)^\prime\dd \bar{n} \right]\right\rbrace\dd n\, .
\label{eq:deltaphi1:sol:N}
\eeq
From this expression, since $\left\langle\xi\left(N\right)\xi\left(N^\prime\right)\right\rangle=\delta\left(N-N^\prime\right)$, it is straightforward to see that\footnote{The $1/2$ factor comes from the relation $\int_{x_1}^{x_2}f(x)\delta(x-x_2)\dd x=f(x_2)/2$, which applies when the Dirac function is centered at a boundary of the integral.}
\beq
\left\langle \xi\delta{\phi^{(1)}} \right\rangle = \frac{H}{4\pi}\, .
\label{eq:stocha:deltaphi1xi}
\eeq
This is why one obtains
\beq
\frac{\dd \left\langle \delta{\phi^{(1)}}^2 \right\rangle }{\dd N}+4\Mp^2\left(\frac{H^\prime}{H}\right)^\prime\left\langle \delta{\phi^{(1)}}^2 \right\rangle=\frac{H^2}{4\pi^2}\, .
\eeq
Since the equation of motion for $\phi_\ucl$ is simply given by $\dd N=-H/(2H^\prime\Mp^2)\dd\phi_\ucl$, one can change the time variable from $N$ to $\phi_\ucl$ and the formal solution of the above equation can be written as
\beq
\left\langle \delta{\phi^{(1)}}^2 \right\rangle = -\frac{1}{8\pi^2\Mp^2}\frac{{H^\prime}^2}{H^2}\int\frac{H^5}{{H^\prime}^3}\dd\phi_\ucl\, .
\label{eq:stocha:deltaphi1Squared}
\eeq
For quadratic inflation where $H=m\phi/(\sqrt{6}\Mp)$, this exactly gives rise to \Eq{eq:stoch:deltaphi1Squared:lfi} while for power-law inflation where\footnote{One can show~\cite{Lucchin:1984yf} that the potential associated with power-law inflation, for which $a(t)\propto t^p$, is given by $V(\phi)\propto \ee^{-\sqrt{2/p}\phi/\Mp}$. Since $H^2=V(\phi)/(3\Mp^2)$ at leading order in slow roll, one obtains the given $H(\phi)$ profile.} $H=H_0\exp\left[-1/\sqrt{2p}(\phi-\phi_0)/\Mp\right]$, one exactly obtains \Eq{eq:stoch:deltaphi1Squared:pli}. Therefore, stochastic and standard field-theoretical approaches to inflation produce the same results for the amount of field fluctuations.\footnote{Here we have established this result at leading order in perturbation theory. However, as shown in \Refs{Starobinsky:1986fx,Starobinsky:1994bd}, the stochastic approach can reproduce QFT results for any finite number of scalar loops and even beyond.}

To emphasize the specificity of $N$ as a preferred time variable, let us repeat the same procedure using the Langevin equation written in terms of $t$, 
\beq
\frac{\dd\tilde{\phi}}{\dd t}=-2\Mp^2H^\prime +\frac{H^{3/2}}{2\pi}\xi\left(t\right)\, .
\eeq
Since this corresponds to a different stochastic process as the one written in terms of $N$, we use again the notation $\tilde{\phi}$ instead of $\phi$. At leading order in the noise, one obtains for $\delta\tilde{\phi}^{(1)}$
\beq
\frac{\dd\delta\tilde{\phi}^{(1)}}{\dd t}+2\Mp^2 H^{\prime\prime}\delta\tilde{\phi}^{(1)}=\frac{H^{3/2}}{2\pi}\xi\left(t\right)\, .
\label{eq:stocha:equadiff:deltatildephi1}
\eeq
Again, multiplying this equation by $\delta\tilde{\phi}^{(1)}$ and taking the stochastic average leads to
\beq
\frac{\dd \left\langle \delta{{}{\tilde{\phi}}^{(1)}}^{2} \right\rangle }{\dd t}+4\Mp^2H^{\prime\prime}\left\langle \delta{{}{\tilde{\phi}}^{(1)}}^{2} \right\rangle=\frac{H^{3/2}}{\pi}\left\langle \xi\left(t\right)\delta\tilde{\phi}^{(1)} \right\rangle\, .
\eeq
In the same manner as before, making use of the formal solution to \Eq{eq:stocha:equadiff:deltatildephi1},
\beq
\delta{\tilde{\phi}^{(1)}}=\exp\left[-2\Mp^2\int_{t_0}^t H^{\prime\prime}\dd u \right]\int_{t_0}^t\left\lbrace \frac{H^{3/2}}{2\pi}\xi\left(u\right)\exp\left[2\Mp^2\int_{t_0}^uH^{\prime\prime}\dd v \right]\right\rbrace\dd u\, ,
\eeq
one can show that $\left\langle \xi\left(t\right)\delta\tilde{\phi}^{(1)} \right\rangle = H^{3/2}/(4\pi)$, so that one needs to solve
\beq
\frac{\dd \left\langle \delta{{}{\tilde{\phi}}^{(1)}}^{2} \right\rangle }{\dd t}+4\Mp^2H^{\prime\prime}\left\langle \delta{{}{\tilde{\phi}}^{(1)}}^{2} \right\rangle=\frac{H^3}{4\pi^2} \, .
\eeq
Making use of the classical trajectory $\dd t=-\dd \phi_\ucl/(2\Mp^2 H^\prime)$, one obtains\footnote{This equation~(\ref{eq:stocha:deltaphi1Squared:t}) also matches Eq.~(13) of \Ref{Martin:2005ir} where perturbative solutions of stochastic inflation are derived when formulated in terms of the cosmic time.}
\beq
\left\langle \delta{{}{\tilde{\phi}}^{(1)}}^{2} \right\rangle = -\frac{{H^\prime}^2}{8\pi^2\Mp^2}\int\frac{H^3}{{H^\prime}^3}\dd\phi_\ucl
\label{eq:stocha:deltaphi1Squared:t}
\eeq
which is clearly different from \Eq{eq:stocha:deltaphi1Squared}.\footnote{As shown in section~\ref{sec:deltaN}, this difference is crucial, since it leads to an incorrect result for the power spectrum of scalar perturbations.} For example, for quadratic inflation, it reduces to $\langle  \delta{{}{\tilde{\phi}}^{(1)}}^{2} \rangle =3(H_0^4-H^4)/(16\pi^2 m^2)$ which does not coincide with \Eq{eq:stoch:deltaphi1Squared:lfi} and for power-law inflation, it reads $\langle  \delta{{}{\tilde{\phi}}^{(1)}}^{2} \rangle = pH^2/(4\pi^2)\ln(H_0/H) $, which does not coincide with \Eq{eq:stoch:deltaphi1Squared:pli}.

Finally and in passing, let us derive the corresponding results for the leading order of the mean fluctuation $\langle \delta\phi\rangle $. Since from \Eq{eq:stocha:equadiff:deltaphi1} it is clear that $\langle \delta\phi^{(1)}\rangle =0$, one has to work out $\langle \delta\phi^{(2)}\rangle $. Expanding $\varphi=\phi_\ucl+\delta\phi^{(1)}+\delta\phi^{(2)}$ in \Eq{eq:Langevin:N:H}, one obtains
\beq
\frac{\dd \delta\phi^{(2)}}{\dd N}+2\Mp^2\left(\frac{H^\prime}{H}\right)^\prime\delta\phi^{(2)} + \Mp^2\left(\frac{H^\prime}{H}\right)^{\prime\prime}{\delta\phi^{(1)}}^2=\frac{H^\prime}{2\pi}\delta\phi^{(1)}\xi\left(N\right)\, .
\eeq
When taking the stochastic average of the above equation, $\langle {\delta\phi^{(1)}}^2\rangle $ is given by \Eq{eq:stocha:deltaphi1Squared} and $\langle \delta\phi^{(1)}\xi\rangle $ is given by \Eq{eq:stocha:deltaphi1xi}, so that one obtains
\beq
\frac{\dd \left\langle\delta\phi^{(2)}\right\rangle}{\dd N}+2\Mp^2\left(\frac{H^\prime}{H}\right)^\prime\left\langle\delta\phi^{(2)}\right\rangle = \frac{1}{8\pi^2}\left(\frac{H^\prime}{H}\right)^{\prime\prime}{\left(\frac{H^\prime}{H}\right)}^ 2\int\frac{H^5}{{H^\prime}^3}\dd\phi+\frac{HH^\prime}{8\pi^2}\, .
\eeq
Using the classical trajectory $\dd \phi_\ucl=-2\Mp^2H^\prime/H\dd N$, this equation can be written in terms of $\phi_\ucl$, and after integration by parts, this gives rise to
\beq
\left\langle\delta\phi^{(2)}\right\rangle = \frac{1}{2}\frac{\left(H^\prime/H\right)^\prime}{H^\prime/H}\left\langle{\delta\phi^{(1)}}^2\right\rangle+\frac{1}{32\Mp^2\pi^2}\frac{H^\prime}{H}\left(\frac{H_0^4}{{H_0^\prime}^2}-\frac{H^4}{{H^\prime}^2}\right)\, ,
\label{eq:stocha:deltaphi2:mean}
\eeq
where $\langle {\delta\phi^{(1)}}^2\rangle $ is given by \Eq{eq:stocha:deltaphi1xi}. For example, when applied to quadratic inflation where $V=m^2\phi^2/2$, one obtains
\beq
\left\langle\delta\phi^{(2)}\right\rangle = \frac{\sqrt{6}}{96\pi^2m\Mp H}\left[\frac{H^6-H_0^6}{H^2}-3\left(H^4-H_0^4\right)\right]\, ,
\eeq
which matches Eq.~(49) of \Ref{Finelli:2008zg}. However, it is again worth noting that one would have obtained a completely different result starting from the Langevin equation written in terms of cosmic time $t$, namely\footnote{This equation,~(\ref{eq:sto:pert:deltaphi2:t}) matches Eq.~(15) of \Ref{Martin:2005ir} where perturbative solutions of the Langevin equation are derived when formulated in terms of the cosmic time.}
\beq
\left\langle\delta\tilde{\phi}^{(2)}\right\rangle = \frac{1}{2}\frac{H^{\prime\prime}}{H^\prime}\left\langle \delta{{}{\tilde{\phi}}^{(1)}}^{2} \right\rangle +\frac{H^\prime}{32\pi^2\Mp^2}\left(\frac{H_0^3}{{H_0^\prime}^2}-\frac{H^3}{{H^\prime}^2}\right)\, .
\label{eq:sto:pert:deltaphi2:t}
\eeq
This obviously differs from \Eq{eq:stocha:deltaphi2:mean}.

To summarize the discussion, different time variables in the Langevin equation lead to different stochastic processes, and the only time variable which allows the stochastic inflation formalism to reproduce QFT calculations is the number of \efolds $N$. One should therefore always work with $N$ when dealing with stochastic inflation.

\bibliographystyle{JHEP}
\bibliography{stochacorr}
\end{document}